%% file: main.tex
\documentclass[aip,reprint,amsmath,amssymb]{revtex4-2}

\usepackage{amsmath,amssymb}
\usepackage{hyperref}
\usepackage{xcolor}
\usepackage{graphicx}
\usepackage{mhchem}

\newcommand{\ket}[1]{\left| #1 \right\rangle}
\newcommand{\bra}[1]{\langle #1 |}
\newcommand{\braket}[2]{\left\langle #1 \middle| #2 \right\rangle}
\newcommand{\mel}[3]{\left\langle #1 \middle| #2 \middle| #3 \right\rangle}
\def\cby{(\textit{S})-3-chloro-1-butyne}
\def\runit{10$^{-44}$ esu$^{2}$ cm$^{2}$}

\begin{document}

\title{Coupled Cluster Simulations of Vibrational Circular Dichroism Spectra}
\author{T. Daniel Crawford}
    \email{crawdad@vt.edu}
    \affiliation{Department of Chemistry, Virginia Tech, Blacksburg, VA 24061, U.S.A.}
\author{Aparna Krishnan}
    \affiliation{Department of Chemistry, Virginia Tech, Blacksburg, VA 24061, U.S.A.}
\author{Brendan M. Shumberger}
    \affiliation{Department of Chemistry, Virginia Tech, Blacksburg, VA 24061, U.S.A.}
\date{\today}

\input{abstract}
\maketitle

\input{intro}
\input{theory}
\input{implementation}
\input{comp}
\input{results}

\input{conclusions}

\section{Supplementary Material} \label{si}

See the supplementary material for the CCSD/cc-pVTZ optimized geometry and
Hessian of \cby; LG-APTs, AATs, and VG-APTs (reference, correlation, and
total); and HF/aug-cc-pCVTZ reference contributions to the LG-APTs, AATs, and
VG-APTs.

\begin{acknowledgments} \label{ack}
This work was supported by the U.S.\ National Science Foundation via grants
CHE-2154753 and OAC-2410880.  The authors are grateful to Advanced Research
Computing at Virginia Tech for providing computational resources that have
contributed to the results reported within the paper.
\end{acknowledgments}

\bibliography{references}
\end{document}


\title{Supplementary Material for: Coupled Cluster Simulations of Vibrational Circular Dichroism Spectra}
\author{T. Daniel Crawford}
    \email{crawdad@vt.edu}
    \affiliation{Department of Chemistry, Virginia Tech, Blacksburg, VA 24061, U.S.A.}
\author{Aparna Krishnan}
    \affiliation{Department of Chemistry, Virginia Tech, Blacksburg, VA 24061, U.S.A.}
\author{Brendan M. Shumberger}
    \affiliation{Department of Chemistry, Virginia Tech, Blacksburg, VA 24061, U.S.A.}
\date{\today}

\maketitle

\section*{Contents}

\begin{itemize}
  \item Table~\ref{tab:si_geom}: CCSD(FC)/cc-pVTZ equilibrium structure of \cby.
  \item Tables~\ref{tab:si_lgapt}--\ref{tab:si_vgapt}: CCSD(FC)/cc-pVTZ LG-APTs, AATs, and VG-APTs
    (reference, correlation, and total), from which the left half of Table~I of the paper is
    computed.
  \item Table~\ref{tab:si_augref}: HF/aug-cc-pCVTZ reference contributions to the LG-APTs, AATs,
    and VG-APTs, from which, together with the correlation contributions of
    Tables~\ref{tab:si_lgapt}--\ref{tab:si_vgapt}, the right half of Table~I is computed.
  \item Tables~\ref{tab:si_hess1}--\ref{tab:si_hess4}: CCSD(FC)/cc-pVTZ Cartesian Hessian.
\end{itemize}

\section*{Conventions}

All quantities are given in atomic units in the center-of-mass principal-axis frame of
Table~\ref{tab:si_geom}, with atoms in the order listed there.  For atom $\lambda$, the APTs
(length or velocity gauge) are tabulated as $P^{\lambda}_{\alpha\gamma} = \partial\mu_\gamma /
\partial R_{\lambda\alpha}$, where the row index $\alpha$ labels the nuclear displacement and the
column index $\gamma$ the dipole component.  The AATs are tabulated as
$\mathrm{Im}\,M^{\lambda}_{\alpha\beta}$ (the tensor is purely imaginary), where the row index
$\alpha$ labels the nuclear displacement and the column index $\beta$ the magnetic-field
component, with the gauge origin $\mathbf{O}$ at the center of mass (the coordinate origin of
Table~\ref{tab:si_geom}).

Each tensor is the sum of nuclear, reference (HF), and correlation (CCSD) contributions.  The
nuclear contributions are not tabulated separately; in atomic units they are
\begin{align}
  \left[P^{\lambda}_{\alpha\gamma}\right]^\mathrm{nuc} &= Z_\lambda\,\delta_{\alpha\gamma}
    \quad \text{(both gauges)}, \\
  \left[\mathrm{Im}\,M^{\lambda}_{\alpha\beta}\right]^\mathrm{nuc} &= \frac{Z_\lambda}{4}
    \sum_\gamma \epsilon_{\alpha\beta\gamma} \left(R_{\lambda\gamma} - O_\gamma\right),
\end{align}
where $Z_\lambda$ and $\mathbf{R}_\lambda$ are the nuclear charge and position.  The ``Total''
columns of Tables~\ref{tab:si_lgapt}--\ref{tab:si_vgapt} include these nuclear terms.  The
tensors entering the right half of Table~I of the paper are
\begin{align}
  \mathbf{T} &= \mathbf{T}^\mathrm{nuc}
             + \mathbf{T}^\mathrm{ref}[\text{HF/aug-cc-pCVTZ}] \notag \\
& \qquad+ \mathbf{T}^\mathrm{corr}[\text{CCSD/cc-pVTZ}],
  \label{eq:si_composite}
\end{align}
with $\mathbf{T}^\mathrm{ref}$ from Table~\ref{tab:si_augref} and $\mathbf{T}^\mathrm{corr}$ from
the ``Correlation'' columns of Tables~\ref{tab:si_lgapt}--\ref{tab:si_vgapt}.

The normal-mode displacements $S_{\lambda\alpha,i}$ are obtained from the Hessian of
Tables~\ref{tab:si_hess1}--\ref{tab:si_hess4} (masses of $^{1}$H, $^{12}$C, and $^{35}$Cl in u;
translations and rotations projected out of the mass-weighted Hessian), and the rotatory
strength of mode $i$ is
\begin{equation}
  R_i = \sum_\beta \mu'_{i\beta}\, m'_{i\beta},
\end{equation}
where
\begin{equation}
  \mu'_{i\beta} = \sum_{\lambda\alpha} P^{\lambda}_{\alpha\beta} S_{\lambda\alpha,i}, \qquad
  m'_{i\beta} = \sum_{\lambda\alpha} \mathrm{Im}\,M^{\lambda}_{\alpha\beta} S_{\lambda\alpha,i}.
\end{equation}
With $S$ in bohr\,u$^{-1/2}$ and the tensors in atomic units, multiplying by $2586.245$ gives
$R_i$ in $10^{-44}$\,esu$^2$\,cm$^2$.  The tabulated values reproduce all frequencies in Table~I of the paper to
within $10^{-3}$\,cm$^{-1}$ and all rotatory strengths to within $10^{-3}$ in these units.

\input{si/si_geometry}
\input{si/si_lgapt}
\input{si/si_aat}
\input{si/si_vgapt}
\input{si/si_augref}
\input{si/si_hessian}

\bibliography{references}

%% file: abstract.tex
\begin{abstract}

We report the first simulations of vibrational circular dichroism (VCD)
spectra at the coupled-cluster level of theory.  We generalize Stephens's
magnetic-field perturbation theory framework for the
magnetic-dipole vibrational transition moment to biorthonormal wave functions.
The resulting atomic axial tensors (AATs) and velocity-gauge atomic polar
tensors collapse to Stephens's expressions for real, Hermitian states.
Adapting these exact-state expressions to coupled-cluster left- and right-hand
biorthonormal wave functions and using our recently developed density-based
formulation yields compact equations that involve only perturbed ground-state
amplitudes, as well as unperturbed and perturbed, unrelaxed one-electron
densities; two-electron-density contributions vanish exactly.  We have
validated the resulting AATs by comparison with our previous configuration
interaction formulation for a two-electron system, size-intensivity tests, a
translational sum rule, and other tests.  Simulated VCD spectra for \cby\ at
the coupled-cluster singles and doubles level with the correlation-consistent
triple-zeta basis set reveal significant discrepancies between length and
velocity gauges.  We find that the addition of large-exponent functions via a
core-valence basis set for the Hartree--Fock reference components yields
substantially improved gauge invariance without the need for gauge-including
atomic orbitals.

\end{abstract}

%% file: intro.tex
\section{Introduction} \label{intro}

The first reports of experimental measurements of vibrational circular
dichroism (VCD) spectra appeared in the early 1970s,\cite{Dudley1972,
Chabay1972, Hsu1973, Schrader1972, Holzwarth1973, Holzwarth1974} and its use
has expanded greatly in the decades since.\cite{Nafie97, Magyarfalvi2011,
Krupova2020, Nafie2020, Merten2024} VCD band intensities of fundamental
($0\rightarrow 1$) transitions in the electronic ground state for a given
vibrational mode $i$ are determined by the rotatory strength,
\begin{equation}
\label{eq:VCD-def}
R(0 \rightarrow 1)_i = \mathrm{Im} \left[ \mel{0}{\vec{\mu}}{1_i} \cdot
\mel{1_i}{\vec{m}}{0} \right],
\end{equation}
where $\vec{\mu}$ and $\vec{m}$ are the electric- and magnetic-dipole
operators. The electric-dipole vibrational transition moment is
straightforwardly computed from the derivative of the ground-state dipole-moment
expectation value with respect to nuclear coordinates [known as the
length-gauge atomic polar tensor (LG-APT)].  The central challenge of the
computational simulation of VCD spectra, however, is the magnetic-dipole
vibrational transition moment, which cannot be computed using the same
approach because the ground-state expectation value of the magnetic-dipole
operator vanishes at all geometries, so the electronic component of the transition moment is zero within
the Born--Oppenheimer approximation.  Several approaches for overcoming this
challenge have been reported over the decades, including the vibronic coupling
theory (VCT) of Craig and Thirunamachandran\cite{Craig1978} and of Nafie and
Freedman,\cite{Nafie1983:VCT} the nuclear-velocity perturbation theory (NVPT)
approach by Nafie,\cite{Nafie83} Stephens's magnetic-field perturbation theory
(MFPT),\cite{Stephens85} the nuclear electromagnetic shielding approach of
Lazzeretti, Zanasi, and Stephens,\cite{Lazzeretti1986} and velocity-dependent
property surfaces by Buckingham, Fowler, and Galwas.\cite{Buckingham1987}

\textit{Ab initio} simulations of VCD spectra were first reported in the 1980s
by Stephens and co-workers at the Hartree--Fock self-consistent field (SCF)
level of theory,\cite{Lowe86:CPL, Lowe86:JACS, Amos1987} based on Stephens's
MFPT.\cite{Stephens85}  This was followed a few years later at the
multiconfigurational SCF (MCSCF) level with the use of gauge-including atomic
orbitals (GIAOs) by Bak et al.,\cite{Bak93, Bak94} and then by Cheeseman,
Frisch, Devlin, and Stephens using density-functional theory (DFT) with
GIAOs.\cite{Cheeseman96} In addition, Yang and Rauk computed VCD intensities
at the second-order M{\o}ller--Plesset (MP2) level within
VCT,\cite{Nafie1983:VCT} using a sum over singly excited configurations
relative to the Hartree--Fock reference.\cite{Yang1994}   In the 2010s,
Scherrer, Vuilleumier, and Sebastiani reported the first DFT-level
implementation of VCD using Nafie's NVPT approach.\cite{Scherrer2013,
Scherrer2016}  In 2022, Luber and co-workers reported DFT-based
implementations of VCD within the CP2K package using NVPT with
velocity-dependent Gaussian basis functions and plane waves, and using MFPT
with GIAOs.\cite{Ditler2022} More recently, our group has reported the first
MP2 and configuration interaction (CID and CISD) implementations of Stephens's
MFPT approach using numerical\cite{Shumberger2024} and analytic-derivative
methods.\cite{Shumberger2025, Shumberger2026, Krishnan2026:density}

In this work, we report the first derivation, implementation, and application
of coupled-cluster theory to the computation of VCD spectra based on an
extension of Stephens's MFPT formulation.  In the next section we discuss in
detail the theoretical foundations of the representation of the
magnetic-dipole vibrational transition moment in a biorthonormal framework and
use this to obtain a generalized expression for the corresponding atomic axial
tensor (AAT) and subsequently for the velocity-gauge atomic polar tensor
(VG-APT).  After demonstrating that the generalized expressions are
independent of the choice of biorthonormal representation, we derive the
coupled-cluster form of the AAT and VG-APT.  Next we discuss our computer
implementation of the tensors and provide details of our efforts to validate
the AAT, in particular.  Finally, we present the first CCSD-level VCD spectra,
using \cby\ as an example.

%% file: theory.tex
\section{Theory} \label{theory}

We extend the pioneering work of Stephens in 1985 on VCD\cite{Stephens85} from
its Hermitian formulation to non-Hermitian, coupled-cluster-type wave
functions.  We begin with the generalization of Stephens's MFPT framework for
the vibrational magnetic-dipole transition moment to biorthonormal pairs of
left- and right-hand Born--Oppenheimer states and then derive from this the
corresponding AAT.  Starting from a biorthonormal expression for the
ground-state dipole derivative, we convert the LG-APT to the VG-APT via the
biorthonormal analogue of the  off-diagonal hypervirial relation.  We find
that, in the biorthonormal framework, both tensors take the form of an
antisymmetric combination of overlaps of left- and right-hand wave-function
derivatives, which reduces to a single overlap in the Hermitian limit.  Next,
we show that a reciprocal rescaling of the right- and left-hand ground-state
wave functions leaves the AAT and VG-APT unchanged due to the biorthonormality
of the wave functions and the inherent antisymmetry of the generalized tensor
expressions.  Finally, we re-formulate the resulting AAT and VG-APT tensors
using the coupled-cluster \textit{Ansatz}.

\subsection{The Vibrational Magnetic-Dipole Transition Moment for Biorthonormal Born--Oppenheimer States}

Let $\ket{\psi_K}$ and $\bra{\tilde\psi_K}$ be right- and left-hand states satisfying
\begin{equation}
H_{\mathrm{el}}\ket{\psi_K} = W_K\ket{\psi_K}
\quad \text{and} \quad \bra{\tilde\psi_K}H_{\mathrm{el}} = \bra{\tilde\psi_K}W_K,
\label{eq:biorth}
\end{equation}
and the biorthonormality condition, 
\begin{equation}
\braket{\tilde\psi_K}{\psi_{K'}} = \delta_{KK'},
\label{eq:bio}
\end{equation}
at every nuclear geometry.  The Born--Oppenheimer products are
$\ket{\Psi_{Kk}} = \ket{\psi_K}\ket{\chi_{Kk}}$ and $\bra{\tilde\Psi_{Kk}} =
\bra{\chi_{Kk}}\bra{\tilde\psi_K}$, with the vibrational functions defined as
eigenfunctions of the sum of the Born--Oppenheimer energy and the nuclear
kinetic-energy operator, $T_\mathrm{N}$ [Eq.~(4) of Ref.~\onlinecite{Stephens85}].
Given the magnetic-dipole operator,
\begin{equation}
\vec{m}^{\mathrm{e}} = - \frac{e}{2m_\mathrm{e}} \sum_n \vec{r}_n \times \vec{p}_n
\end{equation}
where $m_\mathrm{e}$ is the mass of an electron and $n$ runs over the electrons, we take
\begin{equation}
\mel{\tilde\psi_G}{\vec{m}^{\mathrm{e}}}{\psi_G} = 0,
\label{eq:assume}
\end{equation}
at all nuclear geometries, as holds identically for real Born--Oppenheimer
wave functions; this is the only assumption made beyond those of
Ref.~\onlinecite{Stephens85}.  For a real, truncated CC wave function
$\mel{\tilde\psi_G}{\vec{m}^{\mathrm{e}}}{\psi_G}$ does not vanish identically
because the left and right states are distinct. Thus, any non-zero value is
not a physical contribution to the magnetic dipole moment, but a non-Hermitian
artifact, and we ignore it from this point.

The right- and left-hand states corrected to first order are
\begin{align}
\ket{\Psi^{\mathrm{cor}}_{Kk}} &= \ket{\Psi_{Kk}}
 + \sum_{K'k'}\ket{\Psi_{K'k'}}\,c_{K'k';Kk},
\label{eq:right}\\
\bra{\tilde\Psi^{\mathrm{cor}}_{Kk}} &= \bra{\tilde\Psi_{Kk}}
 + \sum_{K'k'}b_{Kk;K'k'}\bra{\tilde\Psi_{K'k'}},
\label{eq:left}
\end{align}
with the coefficients
\begin{align}
c_{K'k';Kk} &= \frac{\mel{\tilde\Psi_{K'k'}}{H}{\Psi_{Kk}}}{E_{Kk}-E_{K'k'}}, \\
b_{Kk;K'k'} &= \frac{\mel{\tilde\Psi_{Kk}}{H}{\Psi_{K'k'}}}{E_{Kk}-E_{K'k'}},
\end{align}
which are obtained from the right- and left-hand first-order Schr\"odinger
equations, respectively.  For $K'\neq K$ only the nuclear kinetic energy
contributes to the matrix elements in the numerators, \textit{i.e.},
\begin{equation}
\mel{\tilde\Psi_{K'k'}}{H}{\Psi_{Kk}}
 = \bra{\chi_{K'k'}}\braket{\tilde\psi_{K'}}{[T_\mathrm{N},\psi_K]}\ket{\chi_{Kk}},
\end{equation}
where
\begin{align}
\braket{\tilde\psi_{K'}}{[T_\mathrm{N},\psi_K]}
 ={}& -\frac{\hbar^2}{2}\sum_\lambda\frac{1}{m_\lambda}
      \braket{\tilde\psi_{K'}}{\vec{\nabla}_{\lambda}^2\psi_K} \notag\\
    & -\hbar^2\sum_\lambda\frac{1}{m_\lambda}
      \braket{\tilde\psi_{K'}}{\vec{\nabla}_{\lambda}\psi_K}\cdot\vec{\nabla}_{\lambda}.
\label{eq:comm}
\end{align}
The vibrational magnetic-dipole transition moment becomes, to first order,
\begin{align}
\mel{\tilde\Psi^{\mathrm{cor}}_{Gg}}{\vec{m}^{\mathrm{e}}}{\Psi^{\mathrm{cor}}_{Ge}}
 ={}& \mel{\tilde\Psi_{Gg}}{\vec{m}^{\mathrm{e}}}{\Psi_{Ge}} \notag\\
    & + \sum_{Kk}\Big[\mel{\tilde\Psi_{Gg}}{\vec{m}^{\mathrm{e}}}{\Psi_{Kk}}\,c_{Kk;Ge} \notag\\
    & \qquad + b_{Gg;Kk}\,\mel{\tilde\Psi_{Kk}}{\vec{m}^{\mathrm{e}}}{\Psi_{Ge}}\Big].
\label{eq:first}
\end{align}
By Eq.~\eqref{eq:assume} the first term and the $K=G$ terms of the sum vanish.
For $K\neq G$ we make the Condon approximation (indicated by the subscript 0) that the magnetic transition
dipoles do not depend on the nuclear coordinates,
\begin{equation*}
        m^\beta_{GK} \equiv \mel{\tilde\psi_G}{(\vec{m}^{\mathrm{e}})_\beta}{\psi_K}_0
\end{equation*}
and
\begin{equation*}
        m^\beta_{KG} \equiv \mel{\tilde\psi_K}{(\vec{m}^{\mathrm{e}})_\beta}{\psi_G}_0,
\end{equation*} 
and we neglect the vibrational contribution to the energy denominators, which
allows the sum over $k$ to be closed (resolution of the identity).  Thus, the
two remaining terms of Eq.~\eqref{eq:first} become
\begin{align}
&\mel{\tilde\Psi^{\mathrm{cor}}_{Gg}}{(\vec{m}^{\mathrm{e}})_\beta}{\Psi^{\mathrm{cor}}_{Ge}}
\notag\\ &\quad = \bra{\chi_{Gg}}\sum_{K\neq
G}\frac{1}{\left(W^0_G-W^0_K\right)}
        \Big[m^\beta_{GK}\braket{\tilde\psi_K}{[T_\mathrm{N},\psi_G]}_0 \notag\\
        &\qquad\quad +\braket{\tilde\psi_G}{[T_\mathrm{N},\psi_K]}_0\,m^\beta_{KG}\Big]
   \ket{\chi_{Ge}}.
\label{eq:twoterms}
\end{align} 
In Stephens's derivation the two terms are combined using
$m^\beta_{GK}=-m^\beta_{KG}$, but for biorthonormal zeroth-order states they
remain distinct.  The kinetic-energy-operator terms of Eq.~\eqref{eq:twoterms}
involve second- and first-derivative matrix elements via Eq.~\eqref{eq:comm},
both of which are constants within the Condon approximation.  The
second-derivative terms contribute in proportion to
$\braket{\chi_{Gg}}{\chi_{Ge}}$ and therefore are zero due to orthogonality of
the vibrational wave functions.  With the derivative couplings
\begin{align}
d^{\lambda\alpha}_{KG} &\equiv \braket{\tilde\psi_K}{\partial_{R_{\lambda\alpha}}\psi_G}_0, \notag\\
d^{\lambda\alpha}_{GK} &\equiv \braket{\tilde\psi_G}{\partial_{R_{\lambda\alpha}}\psi_K}_0,
\end{align}
Eq.~\eqref{eq:twoterms} becomes
\begin{align}
&\mel{\tilde\Psi^{\mathrm{cor}}_{Gg}}{(\vec{m}^{\mathrm{e}})_\beta}{\Psi^{\mathrm{cor}}_{Ge}} \notag\\
&\quad = -\hbar^2\sum_{\lambda\alpha}\frac{1}{m_\lambda}\sum_{K\neq G}
        \frac{m^\beta_{GK}\,d^{\lambda\alpha}_{KG}+d^{\lambda\alpha}_{GK}\,m^\beta_{KG}}{\left(W^0_G-W^0_K\right)}
   \notag\\
&\qquad\quad \times\mel{\chi_{Gg}}{\partial_{R_{\lambda\alpha}}}{\chi_{Ge}} ,
\label{eq:S22}
\end{align}
the analogue of Eq.~(22) of Ref.~\onlinecite{Stephens85}.

Differentiating the right-hand eigenvalue equation of Eq.~\eqref{eq:biorth} with
respect to $R_{\lambda\alpha}$ and projecting onto a different left-hand state gives the
derivative couplings,
\begin{align}
d^{\lambda\alpha}_{KG} &= \frac{g^{\lambda\alpha}_{KG}}{\left(W^0_G-W^0_K\right)}, \notag\\
d^{\lambda\alpha}_{GK} &= -\frac{g^{\lambda\alpha}_{GK}}{\left(W^0_G-W^0_K\right)},
\label{eq:couplings}
\end{align}
where
\begin{equation*}
g^{\lambda\alpha}_{KG} \equiv \mel{\tilde\psi_K}{\partial_{R_{\lambda\alpha}}H_{\mathrm{el}}}{\psi_G}_0
\end{equation*}
and
\begin{equation*}
g^{\lambda\alpha}_{GK} \equiv \mel{\tilde\psi_G}{\partial_{R_{\lambda\alpha}}H_{\mathrm{el}}}{\psi_K}_0.  
\end{equation*}
The vibrational matrix element is the same as Ref.~\onlinecite{Stephens85},
\begin{align}
\mel{\chi_{Gg}}{\partial_{R_{\lambda\alpha}}}{\chi_{Ge}}
 ={}& \frac{m_\lambda}{\hbar^2}\left(E_{Ge}-E_{Gg}\right) \notag\\
    & \times\mel{\chi_{Gg}}{\left(R_{\lambda\alpha}-R^0_{\lambda\alpha}\right)}{\chi_{Ge}} .
\end{align}
Substituting into Eq.~\eqref{eq:S22},
\begin{align}
&\mel{\tilde\Psi^{\mathrm{cor}}_{Gg}}{(\vec{m}^{\mathrm{e}})_\beta}{\Psi^{\mathrm{cor}}_{Ge}}
 = -\left(E_{Ge}-E_{Gg}\right) \notag\\
&\qquad \times\sum_{\lambda\alpha}\sum_{K\neq G}
        \frac{m^\beta_{GK}\,g^{\lambda\alpha}_{KG}-g^{\lambda\alpha}_{GK}\,m^\beta_{KG}}{\left(W^0_G-W^0_K\right)^2}
   \notag\\
&\qquad \times\mel{\chi_{Gg}}{\left(R_{\lambda\alpha}-R^0_{\lambda\alpha}\right)}{\chi_{Ge}}.
\label{eq:S26}
\end{align}
For real Hermitian states
$m^\beta_{GK}=-m^\beta_{KG}$ and $g^{\lambda\alpha}_{KG}=g^{\lambda\alpha}_{GK}$, and Eq.~\eqref{eq:S26} reduces to
Stephens's Eq.~(26).

\subsection{The Atomic Axial Tensor}

The first-order responses of the ground state to a nuclear displacement and,
with $H'=-\vec{m}^{\mathrm{e}} \cdot \vec{H}$, to a magnetic field are
\begin{align}
\ket{\partial_{R_{\lambda\alpha}}\psi_G} &= \sum_{K\neq G}\ket{\psi_K}\frac{g^{\lambda\alpha}_{KG}}{\left(W^0_G-W^0_K\right)},
\notag\\
\bra{\partial_{R_{\lambda\alpha}}\tilde\psi_G} &= \sum_{K\neq G}\frac{g^{\lambda\alpha}_{GK}}{\left(W^0_G-W^0_K\right)}\bra{\tilde\psi_K},
\label{eq:dR}\\
\ket{\partial_{H_\beta}\psi_G} &= -\sum_{K\neq G}\ket{\psi_K}\frac{m^{\beta}_{KG}}{\left(W^0_G-W^0_K\right)},
\notag\\
\bra{\partial_{H_\beta}\tilde\psi_G} &= -\sum_{K\neq G}\frac{m^{\beta}_{GK}}{\left(W^0_G-W^0_K\right)}\bra{\tilde\psi_K},
\label{eq:dH}
\end{align}
and their overlaps are
\begin{align}
\braket{\partial_{R_{\lambda\alpha}}\tilde\psi_G}{\partial_{H_\beta}\psi_G}
 &= -\sum_{K\neq G}\frac{g^{\lambda\alpha}_{GK}\,m^{\beta}_{KG}}{\left(W^0_G-W^0_K\right)^2}, \notag\\
\braket{\partial_{H_\beta}\tilde\psi_G}{\partial_{R_{\lambda\alpha}}\psi_G}
 &= -\sum_{K\neq G}\frac{m^{\beta}_{GK}\,g^{\lambda\alpha}_{KG}}{\left(W^0_G-W^0_K\right)^2},
\label{eq:overlaps}
\end{align}
so that Eq.~\eqref{eq:S26} becomes
\begin{align}
&\mel{\tilde\Psi^{\mathrm{cor}}_{Gg}}{(\vec{m}^{\mathrm{e}})_\beta}{\Psi^{\mathrm{cor}}_{Ge}}
 = -\left(E_{Ge}-E_{Gg}\right) \notag\\
&\qquad \times\sum_{\lambda\alpha}
   \Big[\braket{\partial_{R_{\lambda\alpha}}\tilde\psi_G}{\partial_{H_\beta}\psi_G}
       -\braket{\partial_{H_\beta}\tilde\psi_G}{\partial_{R_{\lambda\alpha}}\psi_G}\Big]_0 \notag\\
&\qquad \times\mel{\chi_{Gg}}{\left(R_{\lambda\alpha}-R^0_{\lambda\alpha}\right)}{\chi_{Ge}} .
\label{eq:S34}
\end{align}
Dropping the subscript $G$ for the electronic ground state, we define the
electronic atomic axial tensor
\begin{equation}
I^{\lambda}_{\alpha\beta}
 = \frac12\Big[\braket{\partial_{R_{\lambda\alpha}}\tilde\psi}{\partial_{H_\beta}\psi}
              -\braket{\partial_{H_\beta}\tilde\psi}{\partial_{R_{\lambda\alpha}}\psi}\Big]_0,
\label{eq:AAT}
\end{equation}
where the subscript $0$ indicates that the derivatives are taken at the reference geometry and $H_\beta = 0$.
With this definition, Eq.~\eqref{eq:S34} has exactly the form of Eq.~(34) of Ref.~\onlinecite{Stephens85},
\begin{align}
&\mel{\tilde\Psi^{\mathrm{cor}}_{Gg}}{(\vec{m}^{\mathrm{e}})_\beta}{\Psi^{\mathrm{cor}}_{Ge}}
 = -2\left(E_{Ge}-E_{Gg}\right) \notag\\
&\qquad \times\sum_{\lambda\alpha}I^{\lambda}_{\alpha\beta}\,\mel{\chi_{Gg}}{\left(R_{\lambda\alpha}-R^0_{\lambda\alpha}\right)}{\chi_{Ge}} ,
\end{align}
and the harmonic-oscillator reduction to Eq.~(39) of Ref.~\onlinecite{Stephens85} is unchanged.  
In addition, for a real,
normalized Hermitian state, $\tilde\psi=\psi$ and
\begin{equation}
\braket{\partial_{H_\beta}\psi}{\partial_{R_{\lambda\alpha}}\psi}=-\braket{\partial_{R_{\lambda\alpha}}\psi}{\partial_{H_\beta}\psi},
\end{equation}
so Eq.~\eqref{eq:AAT} reduces to Stephens's expression,
$I^{\lambda}_{\alpha\beta}=\braket{\partial_{R_{\lambda\alpha}}\psi}{\partial_{H_\beta}\psi}_0$.

\subsection{The Velocity-Gauge Atomic Polar Tensor}

The electronic contribution to the LG-APT,
$\left[E^\lambda_{\alpha\gamma}\right]^\text{LG}$, may be written as the
derivative of the ground-state expectation value of the electric-dipole
operator,
\begin{equation}
\vec{\mu}^{\mathrm{e}} = -e\sum_n \vec{r}_n.
\end{equation}
Adapting the formulation of Amos, Jalkanen, and
Stephens,\cite{Amos1987, Amos1988} to the biorthonormal representation,
\begin{align}
&\left[E^\lambda_{\alpha\gamma}\right]^\text{LG} =
        \left[\partial_{R_{\lambda\alpha}}
        \mel{\tilde\psi_G}{(\vec{\mu}^{\mathrm{e}})_\gamma}{\psi_G}\right]_{\!0} \notag \\
& \qquad = 
        \mel{\partial_{R_{\lambda\alpha}}\tilde\psi_G}{(\vec{\mu}^{\mathrm{e}})_\gamma}{\psi_G}_{\!0}
        +
        \mel{\tilde\psi_G}{(\vec{\mu}^{\mathrm{e}})_\gamma}{\partial_{R_{\lambda\alpha}}\psi_G}_{\!0}.
\end{align}
Given the expansions of the derivatives of the right- and left-hand wave functions in
Eq.~\eqref{eq:dR} and defining the transition moments,
\begin{equation*}
        \mu^\gamma_{KG}\equiv\mel{\tilde\psi_K}{(\vec{\mu}^{\mathrm{e}})_\gamma}{\psi_G}_0
\end{equation*}
and
\begin{equation*}
        \mu^\gamma_{GK}\equiv\mel{\tilde\psi_G}{(\vec{\mu}^{\mathrm{e}})_\gamma}{\psi_K}_0,
\end{equation*}
the dipole derivative becomes
\begin{equation}
\left[E^\lambda_{\alpha\gamma}\right]^\text{LG} = 
\sum_{K\neq G}
        \frac{g^{\lambda\alpha}_{GK}\,\mu^\gamma_{KG}+\mu^\gamma_{GK}\,g^{\lambda\alpha}_{KG}}{\left(W^0_G-W^0_K\right)}.
\label{eq:dipder}
\end{equation}
The off-diagonal hypervirial relation applies to the left- and right-hand states separately as
\begin{equation}
        \mel{\tilde\psi_K}{\vec{\pi}^{\mathrm{e}}}{\psi_G} = \frac{im_\mathrm{e}}{\hbar} \left(W_K-W_G\right)
        \mel{\tilde\psi_K}{\sum_n \vec{r}_n}{\psi_G}
\end{equation}
and
\begin{equation}
        \mel{\tilde\psi_G}{\vec{\pi}^{\mathrm{e}}}{\psi_K} = \frac{im_\mathrm{e}}{\hbar} \left(W_G-W_K\right)
        \mel{\tilde\psi_G}{\sum_n \vec{r}_n}{\psi_K},
\end{equation}
where the total electronic linear-momentum operator is
\begin{equation}
        \vec{\pi}^{\mathrm{e}} = \sum_n \vec{p}_n.
\end{equation}
Defining the transition moments
\begin{equation*}
        \pi^\gamma_{KG}\equiv\mel{\tilde\psi_K}{(\vec{\pi}^{\mathrm{e}})_\gamma}{\psi_G}_0
\end{equation*}
and
\begin{equation*}
        \pi^\gamma_{GK}\equiv\mel{\tilde\psi_G}{(\vec{\pi}^{\mathrm{e}})_\gamma}{\psi_K}_0,
\end{equation*}
the two transition dipole moments become
\begin{equation}
        \mu^\gamma_{KG} = -\frac{ie\hbar}{m_\mathrm{e}} \frac{\pi^\gamma_{KG}}{\left(W_G^0-W_K^0\right)}
\end{equation}
and
\begin{equation}
        \mu^\gamma_{GK} = +\frac{ie\hbar}{m_\mathrm{e}} \frac{\pi^\gamma_{GK}}{\left(W_G^0-W_K^0\right)}.
\end{equation}
Inserting these into Eq.~\eqref{eq:dipder} yields the electronic part of the VG-APT, 
\begin{equation}
\left[E^\lambda_{\alpha\gamma}\right]^\text{VG}
= -\frac{ie\hbar}{m_\mathrm{e}} \sum_{K\neq G} \frac{g^{\lambda\alpha}_{GK} \pi^\gamma_{KG} - \pi^\gamma_{GK}
        g^{\lambda\alpha}_{KG}}{\left(W^0_G-W^0_K\right)^2}.
\label{eq:sos_vgapt}
\end{equation}
Expanding the derivatives of the right- and left-hand wave functions to first order with
respect to a spatially uniform vector potential such that $H' = \vec{A} \cdot \vec{\pi}^{\mathrm{e}}$ gives
\begin{equation}
        \ket{\partial_{A_\gamma} \psi_G} = \sum_{K\neq G} \ket{\psi_K}
        \frac{\pi^\gamma_{KG}}{\left(W^0_G-W^0_K\right)}
\end{equation}
and
\begin{equation}
        \bra{\partial_{A_\gamma} \tilde\psi_G} = \sum_{K\neq G} 
        \frac{\pi^\gamma_{GK}}{\left(W^0_G-W^0_K\right)} \bra{\tilde\psi_K},
\end{equation}
respectively.  
Inserting these expressions and the expressions in
Eq.~\eqref{eq:dR} into Eq.~\eqref{eq:sos_vgapt} 
gives the electronic part of the VG-APT in the biorthonormal representation,
\begin{equation}
\left[E^\lambda_{\alpha\gamma}\right]^\text{VG} = -\frac{2ie\hbar}{m_\mathrm{e}} V^\lambda_{\alpha\gamma}
\end{equation}
where
\begin{equation}
        V^\lambda_{\alpha\gamma} = \frac12\Big[
         \braket{\partial_{R_{\lambda\alpha}}\tilde\psi}{\partial_{A_\gamma}\psi}
              -\braket{\partial_{A_\gamma}\tilde\psi}{\partial_{R_{\lambda\alpha}}\psi}
              \Big]_0,
\label{eq:vgapt}
\end{equation}
where we have again adopted the convention of $\psi = \psi_G$ for simplicity.
This equation has the same form as the AAT in Eq.~\eqref{eq:AAT} with the
replacement of the magnetic-field component, $H_\beta$, with the vector
potential component, $A_\gamma$.

\subsection{Normalization}

Biorthonormality fixes the electronic states only up to a rescaling
$\psi\to c\,\psi$, $\tilde\psi\to\tilde\psi/c$, where $c$ may be complex and may depend on the
geometry and the field.  Taking $x$ and $y$ to be distinct perturbations,
such a rescaling affects the ket and bra wave-function derivatives as
\begin{align}
\ket{\partial_y\psi} &\to c\left(\ket{\partial_y\psi}+(\partial_y\ln c)\ket{\psi}\right), \notag\\
\bra{\partial_x\tilde\psi} &\to c^{-1}\left(\bra{\partial_x\tilde\psi}-(\partial_x\ln c)\bra{\tilde\psi}\right).
\end{align}
This leads to a change in the derivative-wave-function overlaps as
\begin{align}
\braket{\partial_x\tilde\psi}{\partial_y\psi} &\to \braket{\partial_x\tilde\psi}{\partial_y\psi} + 
\braket{\partial_x\tilde\psi}{\psi}(\partial_y\ln c) \notag \\
& \hspace{-0.5cm} - \braket{\tilde\psi}{\partial_y\psi}(\partial_x\ln c) -
\braket{\tilde\psi}{\psi}(\partial_x\ln c)(\partial_y\ln c).
\end{align}
For the antisymmetric forms in Eqs.~\eqref{eq:AAT} and \eqref{eq:vgapt}, the
scaling leads to a change in the AAT/VG-APT of
\begin{align}
&\tfrac12 (\partial_y\ln c)\left[ \braket{\partial_x\tilde\psi}{\psi} + \braket{\tilde\psi}{\partial_x \psi}\right]
\notag\\
&- \tfrac12 (\partial_x\ln c)\left[ \braket{\partial_y\tilde\psi}{\psi} + \braket{\tilde\psi}{\partial_y \psi}\right] =
0,
\end{align}
because each of the bracketed quantities is a derivative of the
biorthonormality condition of Eq.~\eqref{eq:bio}, which holds at all
geometries and field strengths.  The AAT and VG-APT are therefore independent
of this normalization convention, and no correction for the intermediate
normalization of the CC wave function is required, unlike for the diagonal
Born--Oppenheimer correction.\cite{Gauss2006}

\subsection{The Coupled-Cluster Form of the AAT and VG-APT}

Following Ref.~\onlinecite{Krishnan2026:density}, consider the overlap of derivatives with respect to two
parameters $x$ and $y$,
\begin{equation}
S_{xy}\equiv\braket{\partial_x\tilde\psi}{\partial_y\psi},
\end{equation}
and decompose each derivative into a part acting on the wave-function amplitudes
and a part acting on the molecular orbitals,
\begin{align}
\partial_x &= \partial^t_x+\partial^\phi_x, \notag\\
\partial^\phi_x &= \sum_{pq}\braket{\phi_p}{\partial_x\phi_q}\,a_p^\dagger a_q, \notag\\
\left(\partial^\phi_x\right)^\dagger
 &= \sum_{pq}\braket{\partial_x\phi_p}{\phi_q}\,a_p^\dagger a_q.
\end{align}
For coupled cluster, the ground state is
\begin{align}
\ket{\psi} &= e^{T}\ket{\Phi_0}, \notag\\
\bra{\tilde\psi} &= \bra{\Phi_0}\left(1+\Lambda\right)e^{-T},
\label{eq:ccstates}
\end{align}
where $T$ is the cluster operator, $\Lambda$ is the de-excitation operator
containing the Lagrangian multipliers, and $\ket{\Phi_0}$ is the reference
determinant, typically taken to be the Hartree--Fock wave function.  The
amplitude derivatives of the CC ket and bra are
\begin{align}
\ket{\partial^t_y\psi} &= T^y e^{T}\ket{\Phi_0}, \notag\\
\bra{\partial^t_x\tilde\psi}
 &= \bra{\Phi_0}\left[\Lambda^x-\left(1+\Lambda\right)T^x\right]e^{-T},
\end{align}
where $T^x$ and $\Lambda^x$ are built from the first derivatives of the $t$
and $\lambda$ amplitudes.  The resulting overlap has four blocks,
\begin{align}
S_{xy} ={}& \braket{\partial^t_x\tilde\psi}{\partial^t_y\psi}
       + \braket{\partial^t_x\tilde\psi}{\partial^\phi_y\psi} \notag\\
     & + \braket{\partial^\phi_x\tilde\psi}{\partial^t_y\psi}
       + \braket{\partial^\phi_x\tilde\psi}{\partial^\phi_y\psi},
\end{align}
and evaluating them as in Ref.~\onlinecite{Krishnan2026:density} gives
\begin{align}
S_{xy} ={}& \mel{\Phi_0}{\left[\Lambda^x-\left(1+\Lambda\right)T^x\right]T^y}{\Phi_0}
 \notag\\
 & + \sum_{pq}D^{(x,0)}_{pq}\braket{\phi_p}{\partial_y\phi_q}
   + \sum_{pq}D^{(0,y)}_{pq}\braket{\partial_x\phi_p}{\phi_q} \notag\\
 & + \sum_{pq}D^{(0,0)}_{pq}\braket{\partial_x\phi_p}{\partial_y\phi_q} \notag\\
 & + \sum_{pqrs}\Gamma^{(0,0)}_{prqs}\braket{\partial_x\phi_p}{\phi_q}\braket{\phi_r}{\partial_y\phi_s},
\label{eq:Sxy}
\end{align}
where the CC densities are
\begin{align}
D^{(0,0)}_{pq} &= \mel{\Phi_0}{\left(1+\Lambda\right)e^{-T}a_p^\dagger a_q e^{T}}{\Phi_0},
\\
\Gamma^{(0,0)}_{prqs} &= \mel{\Phi_0}{\left(1+\Lambda\right)e^{-T}
   a_p^\dagger a_r^\dagger a_s a_q e^{T}}{\Phi_0}, \\
D^{(x,0)}_{pq} &= \mel{\Phi_0}{\left[\Lambda^x-\left(1+\Lambda\right)T^x\right]
   e^{-T}a_p^\dagger a_q e^{T}}{\Phi_0},
\\
D^{(0,y)}_{pq} &= \mel{\Phi_0}{\left(1+\Lambda\right)e^{-T}a_p^\dagger a_q e^{T}\,T^y}{\Phi_0}.
\end{align}
If we take the antisymmetric combination of the wave-function-derivative
overlaps as required for Eqs.~\eqref{eq:AAT} and \eqref{eq:vgapt}, \textit{viz.}, $S_{xy} - S_{yx}$,
we may simplify each block.

For the amplitude-amplitude block, we note that $T^x$ and $T^y$ are both pure excitation
operators and therefore commute. Thus, 
\begin{align}
&\braket{\partial^t_x\tilde\psi}{\partial^t_y\psi} - \braket{\partial^t_y\tilde\psi}{\partial^t_x\psi} \notag\\
& =\mel{\Phi_0}{\left(\Lambda^x T^y - \Lambda^y T^x\right)}{\Phi_0} \notag \\
& =\sum_{\mu} \left( \lambda^x_\mu t^y_\mu - \lambda^y_\mu t^x_\mu\right),
\end{align}
where the summation is over excitations.

For the amplitude-orbital blocks, recognizing that
$\braket{\partial_x\phi_p}{\phi_q}=-\braket{\phi_p}{\partial_x\phi_q}$,
the half-derivative densities, $D^{(x,0)}_{pq}$ and $D^{(0,x)}_{pq}$, for the same perturbation from each
overlap combine. For example,
\begin{align}
&\braket{\partial^t_x\tilde\psi}{\partial^\phi_y\psi} - \braket{\partial^\phi_y\tilde\psi}{\partial^t_x\psi}
\notag\\
  &\quad = \sum_{pq} \left( D^{(x,0)}_{pq} \braket{\phi_p}{\partial_y\phi_q} -
D^{(0,x)}_{pq} \braket{\partial_y\phi_p}{\phi_q} \right) \notag \\
  &\quad = \sum_{pq} \left( D^{(x,0)}_{pq} + D^{(0,x)}_{pq}\right) \braket{\phi_p}{\partial_y\phi_q} \notag \\
  &\quad = \sum_{pq} D^x_{pq} \braket{\phi_p}{\partial_y\phi_q},
\end{align}
where $D^x_{pq} = D^{(x,0)}_{pq} + D^{(0,x)}_{pq}$ is the first derivative of the unrelaxed one-electron density
with respect to the parameter $x$.

The orbital-orbital block includes two components using the unrelaxed
one-electron and two-electron densities. In the antisymmetric sum, the
latter is
\begin{align}
&\braket{\partial^\phi_x\tilde\psi}{\partial^\phi_y\psi}
 - \braket{\partial^\phi_y\tilde\psi}{\partial^\phi_x\psi} \notag\\
&\quad = \sum_{pqrs}\Gamma^{(0,0)}_{prqs}
  \Big( \braket{\partial_x\phi_p}{\phi_q}\braket{\phi_r}{\partial_y\phi_s} \notag\\
&\qquad\quad - \braket{\partial_y\phi_p}{\phi_q}\braket{\phi_r}{\partial_x\phi_s} \Big) = 0.
\end{align}
The final equality holds because, in the second term, moving both derivatives
to the other side of their overlaps using $\braket{\partial_y\phi_p}{\phi_q} =
- \braket{\phi_p}{\partial_y\phi_q}$ and relabelling $p\leftrightarrow r$,
$q\leftrightarrow s$ reveals that the two terms in parentheses are identical
but of opposite signs because
$\Gamma^{(0,0)}_{rpsq}=\Gamma^{(0,0)}_{prqs}$ in the spin-orbital
representation.

Taking into consideration each of the three simplifications above and setting $x =
R_{\lambda\alpha}$ and $y = H_\beta$, the final form of the electronic
contribution to the CC AAT is
\begin{align}
2\,I^{\lambda}_{\alpha\beta}
        =& \sum_\mu \left(
  \lambda^{R_{\lambda\alpha}}_\mu t^{H_\beta}_\mu 
- \lambda^{H_\beta}_\mu t^{R_{\lambda\alpha}}_\mu
        \right) \notag\\
& + \sum_{pq} D^{R_{\lambda\alpha}}_{pq} \braket{\phi_p}{\partial_{H_\beta}\phi_q} \notag\\
& - \sum_{pq} D^{H_\beta}_{pq} \braket{\phi_p}{\partial_{R_{\lambda\alpha}}\phi_q} \notag\\
& + \sum_{pq}D^{(0,0)}_{pq} \Big(
   \braket{\partial_{R_{\lambda\alpha}}\phi_p}{\partial_{H_\beta}\phi_q} \notag\\
& \qquad -\braket{\partial_{H_\beta}\phi_p}{\partial_{R_{\lambda\alpha}}\phi_q}\Big),
\label{eq:cc_aat}
\end{align}
and that for the VG-APT is
\begin{align}
2 V^{\lambda}_{\alpha\gamma}
        =& \sum_\mu \left(
  \lambda^{R_{\lambda\alpha}}_\mu t^{A_\gamma}_\mu
- \lambda^{A_\gamma}_\mu t^{R_{\lambda\alpha}}_\mu
        \right) \notag\\
& + \sum_{pq} D^{R_{\lambda\alpha}}_{pq} \braket{\phi_p}{\partial_{A_\gamma}\phi_q} \notag\\
& - \sum_{pq} D^{A_\gamma}_{pq} \braket{\phi_p}{\partial_{R_{\lambda\alpha}}\phi_q} \notag\\
& + \sum_{pq}D^{(0,0)}_{pq} \Big(
   \braket{\partial_{R_{\lambda\alpha}}\phi_p}{\partial_{A_\gamma}\phi_q} \notag\\
& \qquad -\braket{\partial_{A_\gamma}\phi_p}{\partial_{R_{\lambda\alpha}}\phi_q}\Big),
\label{eq:cc_apt}
\end{align}
where all 
quantities are evaluated at the reference geometry and zero field ($\vec{H} = \vec{A} = 0$).  As in the
CI case, no two-electron density appears.  For a real CI-type wave function
Eq.~\eqref{eq:cc_aat} reduces to twice Eq.~(27) of Ref.~\onlinecite{Krishnan2026:density}.

%% file: implementation.tex
\section{Implementation and Validation} \label{impl}

The analytic expressions for the electronic contributions to the CCSD AAT and
VG-APT were implemented in \textsc{PyCC}\cite{pycc} based on
Eqs.~\eqref{eq:cc_aat} and \eqref{eq:cc_apt}.  Including the nuclear contribution for each tensor yields the total
AAT
\begin{equation}
M^\lambda_{\alpha\beta} = I^\lambda_{\alpha\beta} + \sum_\gamma \epsilon_{\alpha\beta\gamma}
R_{\lambda\gamma}^0 \frac{i Z_\lambda e}{4 \hbar},
\end{equation}
and the total VG-APT
\begin{equation}
\left[P^\lambda_{\alpha\gamma}\right]^\text{VG} = \left[E^\lambda_{\alpha\gamma}\right]^\text{VG} + Z_\lambda e
\delta_{\alpha\gamma}.
\end{equation}
(The nuclear contribution is the same for both the LG-APT and VG-APT.)
The \textsc{PyCC} infrastructure permits the reuse of the derivatives of the
two-electron integrals and perturbed $T$ and $\Lambda$ amplitudes to avoid
expensive recomputation.  Thus, if the molecular Hessian has been computed,
the AAT and VG-APT require only additional responses with
respect to the magnetic-field or vector-potential components, respectively.

We validated the AAT implementation in five ways. First, we compared the elements of
the CCSD AAT to those of the CISD AAT for the \ce{H4}$^{2+}$ ion (for which CCSD and
CISD are equivalent to full CI) using STO-3G, 6-31G, 6-31G**, and cc-pVDZ
basis sets.  The geometry was chosen to be similar to that of hydrogen
peroxide but with displaced H$-$H bond distances around 1.1 \AA\ and a dihedral
angle of 114$^\circ$ for overall $C_1$ symmetry such that all 36 AAT
elements are non-zero.  Our CISD AAT has been validated against both
finite-difference\cite{Shumberger2024, Shumberger2026} and analytic-derivative
formulations,\cite{Shumberger2026} and every element of the CCSD AAT matches the CISD AAT to at least
$10^{-12}$. 

Second, we tested the size-intensivity of the CCSD AAT by placing a hydrogen
peroxide molecule with displaced O$-$H bonds ($C_1$ symmetry) in a fixed frame
with the coordinate and gauge origins at the O$-$O bond midpoint.  We then
placed a helium atom at distances of 3 and 15 $a_0$ away from the coordinate
origin and compared the components of the CCSD AAT associated with the
\ce{H2O2} molecule only with those computed in the absence of the He atom.
With the 6-31G basis set (for which He has no polarizability), at 3 $a_0$ the
largest absolute deviation element-wise is $2.55 \times 10^{-2}$ a.u., whereas
at 15 $a_0$ the deviation falls to $1.68 \times 10^{-14}$ a.u. In contrast,
the corresponding deviations for CISD, which is not size-intensive, are
$1.71\times 10^{-2}$ a.u.\ and $1.91\times 10^{-3}$ a.u., with the latter unchanged even
at an He distance of 40 $a_0$.

Third, we applied the translational sum rule,
\begin{equation}
\sum_\lambda I^\lambda_{\alpha\beta} = \frac{i}{2\hbar} \partial_{H_\beta} \langle \left(\vec{\pi}^{\mathrm{e}}\right)_\alpha \rangle,
\label{eq:sum_rule}
\end{equation}
in which the derivative with respect to the magnetic field on the right-hand
side includes full amplitude and orbital response at zero field.
This expression is related to rules discussed previously in the
literature.\cite{Lazzeretti1985, Lazzeretti1986, Buckingham1987, Stephens1990,
Peters2022, Steinmetz2025} In particular, this equation connects the
translational sum rule for the electronic part of the AAT by Stephens et al.\
[Eq.~(32) of Ref.~\onlinecite{Stephens1990}] with the static linear-response
expression for the total linear momentum and the angular momentum of
Lazzeretti and Zanasi [Eq.~(71) of Ref.~\onlinecite{Lazzeretti1985}].  This
expression requires that the basis functions depend on position only through
$\vec{r} - \vec{R}_\lambda$ and are field-independent (non-GIAO), that the
gauge origin does not change as nuclei are displaced during the calculation, and that the methods used
are translationally invariant.  Thus, this rule provides a
translational-invariance check of the nuclear-coordinate-derivative components
of our implementation, as well as of the assembly of Eq.~\eqref{eq:cc_aat}.  Evaluating the sum rule for the same \ce{H2O2}
test case as used for the size-intensivity test with the 6-31G basis yields agreement to within $10^{-13}$ a.u.\ between the left- and right-hand
sides of Eq.~\eqref{eq:sum_rule} with and without frozen core orbitals.

In addition, we note that, if we replace the nuclear-coordinate perturbation
with an electric-dipole perturbation, the first term in Eq.~\eqref{eq:cc_aat}
is also the zero-frequency first derivative of the linear response function
for electric and magnetic dipoles (with $F_\alpha$ and $H_\beta$ corresponding to the fields), \textit{i.e.}
\begin{equation}
\sum_\mu \left( \lambda_\mu^{F_\alpha} t_\mu^{H_\beta} - \lambda_\mu^{H_\beta} t_\mu^{F_\alpha} \right) = i \left. \partial_\omega
G'_{\alpha\beta}(\omega)\right|_0,
\end{equation}
assuming no orbital response to the perturbations.  Based on this equation, the amplitude-amplitude block was
compared with the zero-frequency slope of the analytic optical-rotation tensor,
$G'(\omega)$, obtained by numerical differentiation with respect to $\omega$
from frequencies ranging from 0.020 to 0.0025 a.u., agreeing to $8.2\times 10^{-11}$
a.u. for the \ce{H2O2} 6-31G test case. Finally, we confirmed that our implementation gives the same results (to
within ca.\ $10^{-13}$ a.u.)\ for the choice of either canonical or
non-canonical perturbed MOs for either the nuclear-coordinate or
magnetic-field perturbations using the same \ce{H2O2} test case.

%% file: comp.tex
\section{Computational details} \label{comp}

The new formulation of the CCSD AAT and VG-APT was used to compute the VCD
spectrum of (\textit{S})-3-chloro-1-butyne.  The molecular equilibrium
structure was optimized at the frozen-core CCSD/cc-pVTZ
level\cite{Dunning89,Woon93} with \textsc{cfour}\cite{Matthews2020:CFOUR}.
(See Table~S1 of the supplementary material.) The nine lowest-lying
occupied orbitals (C 1s and Cl 1s2s2p) were frozen in all correlated
calculations. The SCF, CC amplitude, and linear-response equations were
converged to $10^{-10}$ in all \textsc{cfour} calculations.  The geometry was
optimized with a convergence criterion of $10^{-7}\,E_\mathrm{h}/a_0$ on the
Cartesian gradient.

All VCD quantities, including the molecular Hessian, the AAT, the LG-APT, and
the VG-APT, were then computed analytically using \textsc{PyCC} at this structure,
using integrals and SCF orbitals from \textsc{Psi4} 1.11\cite{Smith20}. The
SCF energy and density were converged to $10^{-12}$, and the CCSD amplitudes,
$\Lambda$ amplitudes, and their derivatives to $10^{-12}$ in the residuals.
The molecule was placed in its center-of-mass principal-axis frame, and the
center of mass was taken as the gauge origin of the AATs.  Each VCD tensor is
the sum of a nuclear contribution, a Hartree--Fock (reference) contribution,
and a CCSD correlation contribution. As a check on the implementation, the
\textsc{PyCC} energy, gradient, harmonic frequencies, and IR intensities were
compared with analytic CCSD/cc-pVTZ frequency calculations from \textsc{cfour}
at the same structure, with harmonic vibrational frequencies agreeing to within
$10^{-4}$~cm$^{-1}$ and LG IR intensities to within $10^{-3}$~km/mol.

The harmonic analysis used the most abundant isotopes ($^{1}$H, $^{12}$C, $^{35}$Cl), with
translations and rotations projected out of the mass-weighted Hessian. For each normal mode $i$,
the electric- and magnetic-dipole transition moments were obtained by contracting the APTs and AATs
with the Cartesian normal-mode displacements. We report three rotatory strengths for each mode:
\begin{itemize}
  \item $R_i^\mathrm{LG}$, from the LG-APT and the AAT;
  \item $R_i^\mathrm{VG}$, from the VG-APT and the AAT;
  \item $R_i^\mathrm{LGOI}$, the origin-independent length-gauge value of
    Ref.~\onlinecite{Shumberger2026:LGOI}, obtained from the singular-value decomposition (SVD) of the mixed
    length/velocity dipole-strength tensor.
\end{itemize}
The three approximately coincide in the complete-basis limit, so their spread
is largely a measure of basis-set incompleteness in the rotatory strengths.  Note, however,
that, while the LGOI rotatory strength can be computed using the mixed length/velocity dipole-strength tensor, we have
recently shown that the same value of $R_i^\mathrm{LGOI}$ may be obtained using the simpler
expression\cite{Krishnan2026:density}
\begin{equation}
R_i^\mathrm{LGOI} = R_i^\mathrm{VG} \sqrt{\frac{D_i^\mathrm{LG}}{D_i^\mathrm{VG}}},
\end{equation}
where $D_i^\mathrm{LG}$ and $D_i^\mathrm{VG}$ are the dipole strengths in the length and velocity gauges, respectively.
Thus, the LGOI rotatory strength is actually proportional to the VG rotatory strength and always carries the same sign.

A second
measure of basis-set incompleteness is the charge-sum rule for the
VG-APT\cite{Stephens1990},
\begin{equation}
        \sum_{\lambda} \left[P^{\lambda}_{\alpha\gamma}\right]^\text{VG} = q\,\delta_{\alpha\gamma},
  \label{eq:sumrule}
\end{equation}
where $q$ is the total molecular charge.
The LG-APTs satisfy the analogous sum rule in any basis, while the VG-APTs
satisfy Eq.~\eqref{eq:sumrule} only in the complete-basis limit, and thus the residual of
Eq.~(\ref{eq:sumrule}) for the VG-APTs measures how incomplete the basis is
for the velocity-gauge properties.

In addition to the CCSD/cc-pVTZ AAT, LG-APT, and VG-APT, we also evaluated a composite in which each
tensor, $\mathbf{T}$, is assembled as
\begin{align}
  \mathbf{T} &= \mathbf{T}^\mathrm{nuc}
             + \mathbf{T}^\mathrm{ref}[\text{HF/aug-cc-pCVTZ}] \notag \\
& \qquad+ \mathbf{T}^\mathrm{corr}[\text{CCSD/cc-pVTZ}].
  \label{eq:composite}
\end{align}
The all-electron HF/aug-cc-pCVTZ\cite{Kendall92,Woon93,Woon95,Peterson2002} reference contributions were computed with
\textsc{PyCC} at exactly the same structure and in the same frame, so that
Eq.~(\ref{eq:composite}) is a term-by-term replacement. The Hessian, and hence the frequencies and
normal modes, is the CCSD/cc-pVTZ one in both cases.

The spectra were simulated from the harmonic, unscaled frequencies $\tilde\nu_i$ as
\begin{equation}
  \Delta\varepsilon(\tilde\nu) = \frac{\tilde\nu}{2.296\times10^{-39}}
  \sum_i R_i\, g(\tilde\nu;\tilde\nu_i),
\end{equation}
where $\Delta\varepsilon$ is in L\,mol$^{-1}$\,cm$^{-1}$, $\tilde\nu$ in cm$^{-1}$, and $R_i$ in
esu$^2$\,cm$^2$. The line shape $g$ is an area-normalized Lorentzian with a full width at half
maximum of 10\,cm$^{-1}$.

%% file: results.tex
\section{Results and Discussion} \label{results}

\begin{figure*}[tb]
  \centering
  \includegraphics[width=\textwidth]{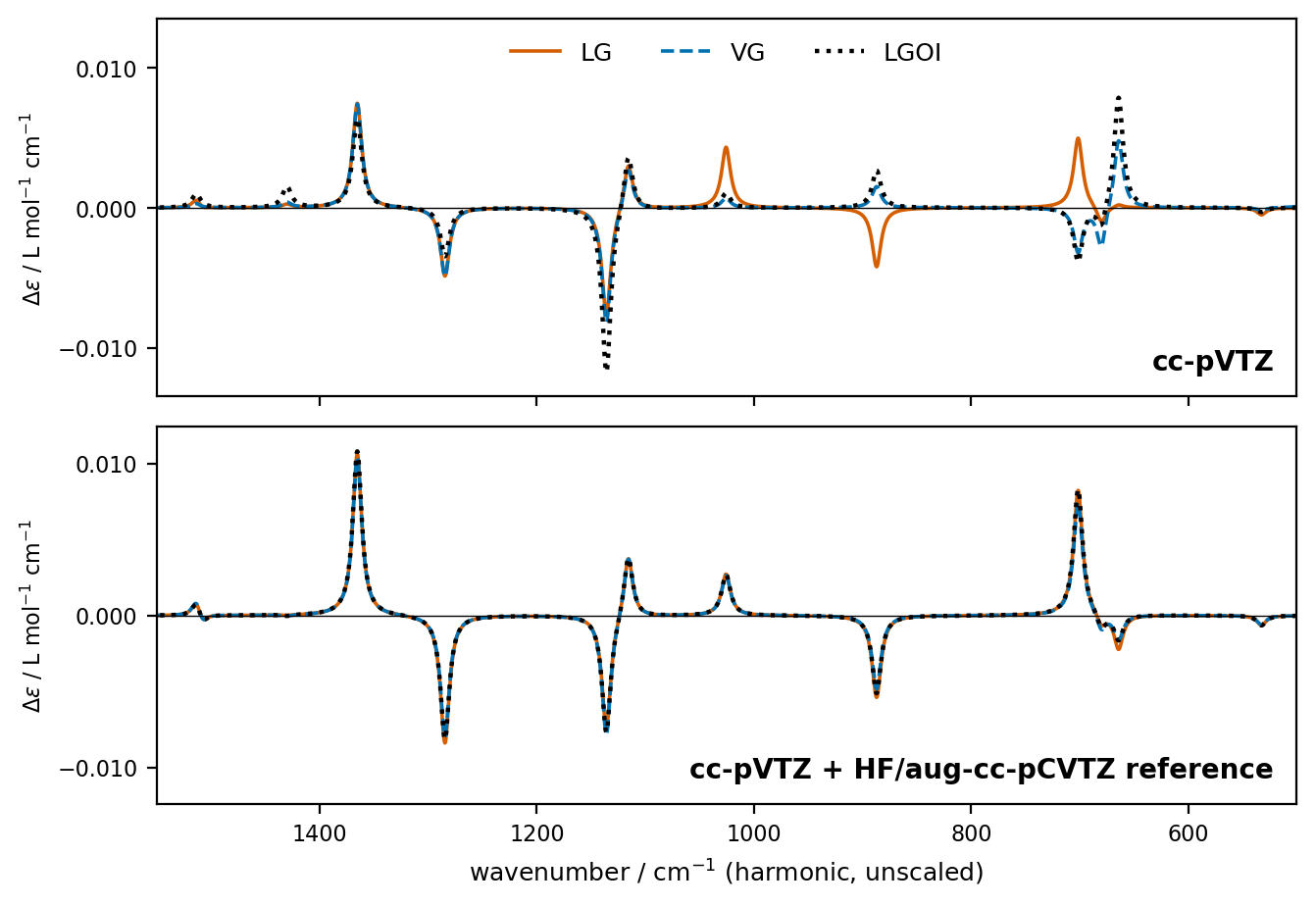}
  \caption{CCSD/cc-pVTZ VCD spectra of (\textit{S})-3-chloro-1-butyne computed with
    \textsc{PyCC}, with rotatory strengths in the length gauge (LG), the velocity gauge (VG), and
    the origin-independent length gauge (LGOI). Top: all tensors in the cc-pVTZ basis. Bottom: the
    reference (HF) contributions to the APTs and AATs replaced by HF/aug-cc-pCVTZ values
    [Eq.~(\ref{eq:composite})]. Harmonic, unscaled frequencies; Lorentzian broadening with a FWHM
    of 10\,cm$^{-1}$; AAT gauge origin at the center of mass. Both panels share the same vertical
    scale.}
  \label{fig:pvtz_gauges}
\end{figure*}

In order to demonstrate the application of the new CCSD-based VCD
implementation, we report in Fig.~\ref{fig:pvtz_gauges} the CCSD/cc-pVTZ VCD
mid-IR spectrum of \cby\ using the three available gauges: LG, VG, and LGOI.
The numerical values of the vibrational frequencies, dipole strengths,
rotatory strengths, and the cosines of the angle between the LG and VG
electric-dipole transition moments are reported in
Table~\ref{tab:cby_pvtz_vcd_augref} for the full range of normal modes.  \cby\
provides a useful example because it exhibits a single conformation and its
mid-IR bands are well separated.\cite{He2004} (See the supplementary material
for the LG-APTs, AATs, and VG-APTs in Tables~S2--S4, and the molecular Hessian
in Tables~S6--S9.)

\input{tables/vcd_pvtz_table_augref}

The upper spectrum in Fig.~\ref{fig:pvtz_gauges} exhibits numerous sign
discrepancies between the LG vs.\ VG and LGOI rotatory strengths, with the
bands at 887.0~cm$^{-1}$ and 701.3~cm$^{-1}$ being the most significant.
Table~\ref{tab:cby_pvtz_vcd_augref} indicates that the values of $\cos{\theta}$
for these two bands are 0.466 and $-0.987$, respectively, indicating that the
LG and VG electric-dipole transition moments are far from alignment ($62^\circ$) for the former
and nearly anti-parallel for the latter.  In addition, while the band at
1025.6~cm$^{-1}$ exhibits no sign discrepancy among the gauges, the LG rotatory strength
is a factor of six larger than its VG counterpart, driven primarily by the
near-perpendicularity of the LG and VG transition-moment vectors ($\theta = 73^\circ$)
and secondarily by the larger LG transition moment.  Overall, six sign
discrepancies occur between the VG/LGOI and LG rotatory strengths for the
CCSD/cc-pVTZ VCD spectra, and the RMS differences between the VG vs.\ LG and the
LGOI vs.\ LG are $13.88 \times$\runit\ and $15.12 \times$\runit, respectively.

The source of the above discrepancies is primarily the incompleteness of the
cc-pVTZ basis set. In general, the LG-APT converges more rapidly to the
complete basis-set (CBS) limit than the AAT or VG-APT, and thus the LG dipole
strengths are expected to be more accurate than the VG values, though the LG
rotatory strengths remain limited by the AAT.  As noted earlier, a measure of
the completeness of the basis set for the VG-APT is given by the charge-sum
rule.  For the CCSD/cc-pVTZ VG-APTs, the largest element of the sum rule is
13.49 a.u., far from the value of zero expected for a neutral molecule,
indicating poor basis-set representation of the VG-APT.  Furthermore, the
nuclear plus Hartree--Fock reference contribution to this value is 13.26 a.u.,
which is approximately 98\% of the total.

Given the above observations, we carried out a second set of computations of
only the Hartree--Fock-level AAT, LG-APT, and VG-APT using the aug-cc-pCVTZ
basis set at the CCSD/cc-pVTZ optimized geometry and used these to replace
their cc-pVTZ counterparts.  (See Table~S5 of the supplementary material.) The
resulting mid-IR VCD spectra are given in the lower frame of
Fig.~\ref{fig:pvtz_gauges} and the right half of
Table~\ref{tab:cby_pvtz_vcd_augref}. This replacement reduces the VG-APT
charge-sum-rule error from 13.49 to 1.09 a.u., leaving the correlation
contribution of 0.23 a.u.\ unchanged (about 20\% of the remaining
discrepancy).  As can be seen in the figure and the table, the agreement among
the gauges is substantially improved with the number of sign discrepancies
reduced to only one, \textit{viz.}\ the lowest-frequency torsion mode for
which the rotatory strengths are all relatively small.  Furthermore, the RMS
spread of the LG vs.\ VG and LG vs.\ LGOI values is also significantly reduced
to $1.49\times$\runit\ and $1.08\times$\runit, respectively.  Finally, the value of $\cos{\theta}$
for every mode above 500~cm$^{-1}$ is now essentially 1.0, with only the
low-frequency modes exhibiting significant deviations from parallel LG and VG
electric-dipole transition moments.

We carried out a more detailed analysis using separate reference calculations
with and without the high-exponent basis functions for each angular-momentum
type ($s$, $p$, $d$, and $f$).  These computations indicate that only the $p$-
and $d$-type functions contribute to the improvement and are approximately
evenly split in their contributions to the total. The $p$ shells carry
essentially all the effect on the carbon atoms and ca.\ 30\% of the effect on
chlorine, and the $d$ shells contribute the remaining 70\% of the effect on
Cl.  Diffuse functions affect the final rotatory strengths roughly 5-10 times
less than the core-valence functions, though they appear to be important for
achieving complete agreement.  In addition, the new basis-set contributions
affect only the AAT and the VG-APT and have no significant impact on the
LG-APT.  Further basis-set analysis is outside the scope of the present work,
but will be the subject of a more systematic study in the near future.

%% file: tables/vcd_pvtz_table_augref.tex

\begin{table*}
\caption{Harmonic vibrational frequencies $\tilde\nu$ (cm$^{-1}$), dipole strengths $D$
($10^{-40}$\,esu$^2$\,cm$^2$), and rotatory strengths $R$ ($10^{-44}$\,esu$^2$\,cm$^2$) of
(\textit{S})-3-chloro-1-butyne at the CCSD(FC)/cc-pVTZ level. Left: all tensors in the cc-pVTZ
basis. Right: reference (HF) contributions to the APTs and AATs from HF/aug-cc-pCVTZ
[Eq.~(\ref{eq:composite})]. LG, VG, and LGOI denote the length gauge, velocity gauge, and
origin-independent length gauge. $\cos\theta$ is the cosine of the angle between the LG and VG
electric-dipole transition moments of each mode. AAT gauge origin at the center of mass. RMS
deviations (same units as $R$) and the number of sign differences are over all 24 modes.}
\label{tab:cby_pvtz_vcd_augref}
\begin{ruledtabular}
\begin{tabular}{rr rrrrrr c rrrrrr}
 & & \multicolumn{6}{c}{cc-pVTZ} & & \multicolumn{6}{c}{cc-pVTZ + HF/aug-cc-pCVTZ reference} \\
\cline{3-8}\cline{10-15}
Mode & $\tilde\nu$ & $D^\mathrm{LG}$ & $D^\mathrm{VG}$ & $\cos\theta$
     & $R^\mathrm{LG}$ & $R^\mathrm{VG}$ & $R^\mathrm{LGOI}$ &
     & $D^\mathrm{LG}$ & $D^\mathrm{VG}$ & $\cos\theta$
     & $R^\mathrm{LG}$ & $R^\mathrm{VG}$ & $R^\mathrm{LGOI}$ \\
\hline
1 & 3491.2 & 66.6 & 36.6 & 1.000 & 0.41 & 0.27 & 0.37 &  & 67.8 & 66.1 & 1.000 & 0.27 & 0.34 & 0.34 \\
2 & 3165.9 & 14.8 & 46.4 & 0.998 & $-$2.40 & $-$7.37 & $-$4.16 &  & 13.1 & 12.0 & 1.000 & $-$2.50 & $-$2.40 & $-$2.51 \\
3 & 3150.9 & 16.0 & 45.4 & 0.998 & 1.21 & 8.46 & 5.03 &  & 14.4 & 13.5 & 1.000 & 1.42 & 1.19 & 1.23 \\
4 & 3128.5 & 1.5 & 6.1 & 0.927 & 0.72 & $-$1.21 & $-$0.60 &  & 0.8 & 0.7 & 0.997 & 0.82 & 0.88 & 0.93 \\
5 & 3070.8 & 14.1 & 27.3 & 0.998 & 1.25 & 0.91 & 0.66 &  & 13.4 & 12.5 & 1.000 & 1.15 & 1.12 & 1.16 \\
6 & 2226.2 & 2.6 & 2.2 & 0.990 & $-$0.32 & 0.01 & 0.01 &  & 1.3 & 1.6 & 1.000 & $-$0.37 & $-$0.47 & $-$0.43 \\
7 & 1513.5 & 4.0 & 0.4 & 0.755 & 1.44 & 0.72 & 2.40 &  & 4.2 & 5.2 & 0.999 & 2.40 & 2.52 & 2.26 \\
8 & 1507.7 & 16.4 & 6.4 & 0.992 & $-$0.44 & $-$0.06 & $-$0.10 &  & 18.0 & 19.7 & 1.000 & $-$1.53 & $-$1.59 & $-$1.53 \\
9 & 1430.0 & 12.3 & 0.7 & 0.715 & 0.59 & 0.92 & 3.78 &  & 13.4 & 15.9 & 1.000 & $-$0.10 & $-$0.23 & $-$0.21 \\
10 & 1365.3 & 27.4 & 37.2 & 0.984 & 19.81 & 19.64 & 16.86 &  & 27.7 & 26.0 & 1.000 & 28.25 & 27.79 & 28.67 \\
11 & 1284.6 & 89.9 & 179.4 & 0.979 & $-$13.76 & $-$13.80 & $-$9.77 &  & 92.7 & 89.2 & 1.000 & $-$23.67 & $-$22.55 & $-$22.99 \\
12 & 1135.8 & 40.0 & 19.4 & 0.862 & $-$25.01 & $-$26.54 & $-$38.07 &  & 39.6 & 40.9 & 1.000 & $-$24.50 & $-$25.58 & $-$25.18 \\
13 & 1115.7 & 48.1 & 26.3 & 0.932 & 11.06 & 10.14 & 13.70 &  & 48.1 & 47.0 & 1.000 & 13.48 & 13.45 & 13.62 \\
14 & 1025.6 & 49.4 & 21.8 & 0.300 & 15.28 & 2.47 & 3.72 &  & 48.2 & 46.3 & 0.999 & 9.63 & 9.32 & 9.51 \\
15 & 887.0 & 73.2 & 23.7 & 0.466 & $-$17.16 & 6.17 & 10.85 &  & 73.1 & 65.3 & 0.999 & $-$21.98 & $-$19.69 & $-$20.83 \\
16 & 701.3 & 285.9 & 186.0 & $-$0.987 & 25.94 & $-$16.28 & $-$20.18 &  & 284.8 & 227.5 & 1.000 & 42.87 & 38.58 & 43.17 \\
17 & 679.9 & 82.3 & 246.1 & 0.977 & $-$6.52 & $-$16.16 & $-$9.34 &  & 81.7 & 98.2 & 1.000 & $-$4.47 & $-$6.04 & $-$5.51 \\
18 & 664.1 & 250.2 & 99.8 & 0.912 & 1.26 & 27.75 & 43.95 &  & 251.2 & 253.2 & 1.000 & $-$12.46 & $-$9.17 & $-$9.13 \\
19 & 532.4 & 59.2 & 30.1 & $-$0.928 & $-$3.39 & $-$1.47 & $-$2.06 &  & 61.0 & 54.7 & 0.999 & $-$4.39 & $-$4.17 & $-$4.40 \\
20 & 473.2 & 17.5 & 1911.4 & $-$0.326 & $-$3.96 & 20.63 & 1.97 &  & 18.4 & 18.8 & 0.746 & $-$5.13 & $-$2.22 & $-$2.19 \\
21 & 322.7 & 18.4 & 2766.1 & $-$0.792 & $-$15.07 & 1.46 & 0.12 &  & 18.6 & 8.7 & 0.597 & 5.29 & 3.18 & 4.64 \\
22 & 272.3 & 4.0 & 513.0 & 0.199 & $-$1.46 & $-$5.95 & $-$0.52 &  & 4.3 & 8.2 & 0.858 & $-$6.13 & $-$6.05 & $-$4.41 \\
23 & 216.7 & 28.3 & 605.1 & 0.122 & 13.09 & 11.63 & 2.52 &  & 31.2 & 34.2 & 0.953 & 2.08 & 2.11 & 2.02 \\
24 & 164.2 & 34.6 & 5889.0 & $-$0.735 & $-$0.27 & $-$19.05 & $-$1.46 &  & 36.8 & 15.8 & 0.570 & 0.47 & $-$0.34 & $-$0.52 \\
\hline
\multicolumn{5}{l}{RMS $(R^\mathrm{VG}-R^\mathrm{LG})$} &  & 13.88 &  &  &  &  &  &  & 1.49 &  \\
\multicolumn{5}{l}{RMS $(R^\mathrm{LGOI}-R^\mathrm{LG})$} &  &  & 15.12 &  &  &  &  &  &  & 1.08 \\
\multicolumn{5}{l}{Sign differences, $R^\mathrm{VG}$ vs.\ $R^\mathrm{LG}$} &  & 6 &  &  &  &  &  &  & 1 &  \\
\end{tabular}
\end{ruledtabular}
\end{table*}

%% file: conclusions.tex
\section{Conclusions} \label{conclusions}

We have developed the first CC-level simulations of VCD spectra.  Using a
biorthonormal-wave-function framework, we have extended Stephens's MFPT
formulation of the magnetic-dipole vibrational transition moment to
accommodate the non-Hermitian structure of CC wave functions, leading to
generalized antisymmetric expressions for the AAT and the VG-APT.  We have
demonstrated that these expressions do not depend on the choice of
biorthonormal representation (unlike the related expressions of the DBOC) and
that they reduce to Stephens's MFPT expressions for real, Hermitian states.
The corresponding CC formulation follows directly from the antisymmetric form,
and our recent density-based framework\cite{Krishnan2026:density} yields compact
equations that depend only on perturbed $\hat{T}$ and $\hat{\Lambda}$ amplitudes,
as well as unperturbed and perturbed (unrelaxed) one-electron densities.
Contributions from the two-electron density vanish exactly, again, unlike for
the DBOC.\cite{Krishnan2026:density}  All components of the VCD-simulation
workflow --- molecular Hessian, AAT, LG-APT, and VG-APT --- have been
implemented in the \textsc{PyCC} package, which is built on top of
\textsc{Psi4}.

We have validated the AAT through (a) comparison to our previous CISD
implementation for chiral \ce{H4}$^{2+}$ for which CCSD and CISD are
equivalent to full CI, (b) check of the size-intensivity of a
\ce{H2O2}$\cdots$\ce{He} supermolecule pair, (c) a translational sum rule
connecting the sum of components of the AAT over nuclei to derivatives of the
electronic linear momentum with respect to an external magnetic field,
(d) comparison of the amplitude-amplitude block against the zero-frequency
slope of the CC optical-rotation tensor, and (e) invariance to the choice of
canonical vs.\ non-canonical perturbed orbitals.

As an initial test of the new implementation, we have computed the
CCSD/cc-pVTZ VCD spectrum of \cby, a conformationally rigid molecule with
well-separated mid-IR bands.  We find significant differences
between the LG and VG/LGOI spectra exist, including multiple sign
discrepancies, which we tie to basis-set incompleteness of the AAT and VG-APT.
Guided by large deviations of the charge-sum rule for the VG-APT, especially
for the reference contributions, we find that combining HF/aug-cc-pCVTZ AAT,
VG-APT, and LG-APT property tensors with their CCSD/cc-pVTZ correlation
counterparts dramatically reduces the gauge discrepancies in the rotatory
strengths and corresponding spectra.  This preliminary result suggests that
such core-valence basis sets may provide an inexpensive strategy for reducing
basis-set effects while still providing gauge-origin invariance.

Future work will involve implementation of GIAOs within the new CC framework,
as well as a systematic study of basis-set effects on the AAT and VG-APT,
including core-valence vs.\ diffuse functions and convergence toward the CBS
limit.  In addition, we plan to implement production-level codes for improved
performance, to compare directly with experimental measurements, and to
develop reduced-scaling/local-CC methods to reach larger molecular systems and
to account for explicit-solvation effects.

%% file: si/si_geometry.tex
\begin{table}[h]
\caption{CCSD(FC)/cc-pVTZ equilibrium structure of \cby\ (bohr) in the center-of-mass principal-axis frame used for all tensors in this document. Atom labels define the atom order of Tables~\ref{tab:si_lgapt}--\ref{tab:si_augref}.}
\label{tab:si_geom}
\begin{ruledtabular}
\begin{tabular}{lrrr}
Atom & \multicolumn{1}{c}{$x$} & \multicolumn{1}{c}{$y$} & \multicolumn{1}{c}{$z$} \\
\hline
C1 & $-$0.0711765739 & $-$0.8235189276 & $-$0.8386716491 \\
C2 & $-$2.4845091385 & 0.3691718897 & $-$0.2186061814 \\
Cl3 & 2.5013936990 & 1.2019291077 & 0.1360305529 \\
C4 & 0.2180607455 & $-$3.4113253083 & 0.3831151962 \\
H5 & 0.1104279040 & $-$0.9660739720 & $-$2.8795740174 \\
C6 & $-$4.5045872036 & 1.2852878365 & 0.2869450360 \\
H7 & $-$6.2750724327 & 2.1143235363 & 0.7357472737 \\
H8 & $-$1.3034526343 & $-$4.6431266638 & $-$0.2578091889 \\
H9 & 2.0261723502 & $-$4.2401303850 & $-$0.1378724443 \\
H10 & 0.1192531757 & $-$3.2445449489 & 2.4301429801 \\
\end{tabular}
\end{ruledtabular}
\end{table}

%% file: si/si_lgapt.tex
\begin{table*}
\caption{CCSD(FC)/cc-pVTZ length-gauge atomic polar tensors (LG-APTs) of \cby. The total includes the nuclear contribution (see text). Rows: nuclear displacement $R_{\lambda\alpha}$; columns: dipole component $\gamma$, i.e., $P^{\lambda}_{\alpha\gamma}$ in atomic units ($e$).}
\label{tab:si_lgapt}
\begin{ruledtabular}
\begin{tabular}{llrrr rrr rrr}
 &  & \multicolumn{3}{c}{Reference (HF)} & \multicolumn{3}{c}{Correlation (CCSD)} & \multicolumn{3}{c}{Total} \\
\cline{3-5}\cline{6-8}\cline{9-11}
Atom & $\alpha$ & \multicolumn{1}{c}{$\gamma=$x} & \multicolumn{1}{c}{$\gamma=$y} & \multicolumn{1}{c}{$\gamma=$z} & \multicolumn{1}{c}{$\gamma=$x} & \multicolumn{1}{c}{$\gamma=$y} & \multicolumn{1}{c}{$\gamma=$z} & \multicolumn{1}{c}{$\gamma=$x} & \multicolumn{1}{c}{$\gamma=$y} & \multicolumn{1}{c}{$\gamma=$z} \\
\hline
C1 & x & $-$4.984593 & 0.235916 & 0.068203 & $-$0.174498 & $-$0.057820 & $-$0.018199 & 0.840909 & 0.178096 & 0.050004 \\
 & y & 0.318067 & $-$5.439540 & 0.076558 & $-$0.075457 & $-$0.057549 & $-$0.027519 & 0.242610 & 0.502911 & 0.049039 \\
 & z & 0.129412 & 0.137092 & $-$5.721907 & $-$0.036118 & $-$0.046465 & $-$0.047861 & 0.093294 & 0.090627 & 0.230233 \\[3pt]
C2 & x & $-$6.066810 & $-$0.081586 & $-$0.036092 & $-$0.021601 & 0.015568 & 0.001114 & $-$0.088411 & $-$0.066018 & $-$0.034978 \\
 & y & $-$0.057480 & $-$6.225804 & $-$0.005978 & 0.023277 & 0.024399 & 0.007343 & $-$0.034202 & $-$0.201405 & 0.001365 \\
 & z & $-$0.047662 & $-$0.002830 & $-$6.158892 & 0.011761 & 0.005831 & 0.020204 & $-$0.035901 & 0.003001 & $-$0.138688 \\[3pt]
Cl3 & x & $-$17.618490 & $-$0.190018 & $-$0.035418 & 0.117368 & 0.050069 & 0.018472 & $-$0.501123 & $-$0.139949 & $-$0.016945 \\
 & y & $-$0.255762 & $-$17.389033 & $-$0.037408 & 0.064687 & 0.064856 & 0.015668 & $-$0.191075 & $-$0.324177 & $-$0.021740 \\
 & z & $-$0.113674 & $-$0.066515 & $-$17.187509 & 0.030589 & 0.020017 & 0.030590 & $-$0.083085 & $-$0.046499 & $-$0.156919 \\[3pt]
C4 & x & $-$6.001445 & $-$0.032964 & 0.002482 & $-$0.031241 & 0.006361 & 0.001738 & $-$0.032686 & $-$0.026604 & 0.004220 \\
 & y & $-$0.049764 & $-$5.917110 & 0.037734 & 0.008158 & $-$0.071199 & $-$0.002043 & $-$0.041606 & 0.011691 & 0.035691 \\
 & z & $-$0.003750 & $-$0.030751 & $-$5.893971 & 0.001854 & 0.011760 & $-$0.041090 & $-$0.001896 & $-$0.018991 & 0.064939 \\[3pt]
H5 & x & $-$1.073684 & $-$0.013447 & $-$0.011780 & 0.013802 & 0.002129 & 0.000841 & $-$0.059882 & $-$0.011318 & $-$0.010940 \\
 & y & $-$0.027382 & $-$1.005525 & $-$0.011724 & 0.004031 & 0.001925 & 0.000105 & $-$0.023351 & $-$0.003600 & $-$0.011619 \\
 & z & $-$0.026038 & $-$0.027941 & $-$1.045654 & 0.008035 & 0.003889 & 0.006458 & $-$0.018003 & $-$0.024051 & $-$0.039196 \\[3pt]
C6 & x & $-$6.448921 & 0.076741 & 0.008346 & 0.078386 & $-$0.015923 & $-$0.001559 & $-$0.370535 & 0.060819 & 0.006787 \\
 & y & 0.095337 & $-$6.231796 & 0.011102 & $-$0.026437 & 0.024612 & $-$0.004859 & 0.068899 & $-$0.207184 & 0.006242 \\
 & z & 0.053184 & $-$0.003169 & $-$6.218244 & $-$0.013392 & $-$0.002087 & 0.023926 & 0.039792 & $-$0.005256 & $-$0.194317 \\[3pt]
H7 & x & $-$0.753594 & $-$0.006167 & $-$0.000548 & $-$0.021799 & $-$0.000389 & $-$0.000381 & 0.224607 & $-$0.006557 & $-$0.000929 \\
 & y & $-$0.001150 & $-$0.771208 & $-$0.004518 & $-$0.000170 & $-$0.021507 & 0.001816 & $-$0.001320 & 0.207285 & $-$0.002702 \\
 & z & $-$0.001244 & $-$0.003035 & $-$0.772260 & 0.000045 & 0.001564 & $-$0.023153 & $-$0.001199 & $-$0.001472 & 0.204587 \\[3pt]
H8 & x & $-$1.033832 & $-$0.097432 & $-$0.053451 & 0.015361 & 0.012368 & 0.008111 & $-$0.018471 & $-$0.085064 & $-$0.045340 \\
 & y & $-$0.102674 & $-$1.051850 & $-$0.016842 & 0.016437 & 0.017023 & 0.005810 & $-$0.086238 & $-$0.034827 & $-$0.011032 \\
 & z & $-$0.041469 & $-$0.016100 & $-$0.953732 & 0.005069 & 0.003325 & 0.003704 & $-$0.036400 & $-$0.012775 & 0.049972 \\[3pt]
H9 & x & $-$1.089849 & 0.100280 & 0.046244 & 0.024178 & $-$0.013332 & $-$0.007928 & $-$0.065671 & 0.086948 & 0.038316 \\
 & y & 0.075679 & $-$1.000291 & $-$0.001929 & $-$0.014283 & 0.011632 & 0.002801 & 0.061396 & 0.011341 & 0.000871 \\
 & z & 0.043546 & $-$0.003618 & $-$0.950298 & $-$0.006416 & 0.003648 & 0.001773 & 0.037130 & 0.000029 & 0.051476 \\[3pt]
H10 & x & $-$0.928782 & 0.008678 & 0.012013 & 0.000045 & 0.000969 & $-$0.002209 & 0.071263 & 0.009647 & 0.009804 \\
 & y & 0.005130 & $-$0.967843 & $-$0.046994 & $-$0.000243 & 0.005808 & 0.000879 & 0.004887 & 0.037965 & $-$0.046116 \\
 & z & 0.007694 & 0.016868 & $-$1.097533 & $-$0.001426 & $-$0.001481 & 0.025448 & 0.006268 & 0.015387 & $-$0.072085 \\
\end{tabular}
\end{ruledtabular}
\end{table*}

%% file: si/si_aat.tex
\begin{table*}
\caption{CCSD(FC)/cc-pVTZ atomic axial tensors (AATs) of \cby. The total includes the nuclear contribution (see text). Rows: nuclear displacement $R_{\lambda\alpha}$; columns: magnetic-field component $\beta$, i.e., $\mathrm{Im}\,M^{\lambda}_{\alpha\beta}$ in atomic units; gauge origin at the center of mass (the coordinate origin of Table~\ref{tab:si_geom}).}
\label{tab:si_aat}
\begin{ruledtabular}
\begin{tabular}{llrrr rrr rrr}
 &  & \multicolumn{3}{c}{Reference (HF)} & \multicolumn{3}{c}{Correlation (CCSD)} & \multicolumn{3}{c}{Total} \\
\cline{3-5}\cline{6-8}\cline{9-11}
Atom & $\alpha$ & \multicolumn{1}{c}{$\beta=$x} & \multicolumn{1}{c}{$\beta=$y} & \multicolumn{1}{c}{$\beta=$z} & \multicolumn{1}{c}{$\beta=$x} & \multicolumn{1}{c}{$\beta=$y} & \multicolumn{1}{c}{$\beta=$z} & \multicolumn{1}{c}{$\beta=$x} & \multicolumn{1}{c}{$\beta=$y} & \multicolumn{1}{c}{$\beta=$z} \\
\hline
C1 & x & 0.058429 & 0.950528 & $-$0.895122 & $-$0.011841 & 0.037334 & $-$0.028058 & 0.046588 & $-$0.270145 & 0.312099 \\
 & y & $-$1.100876 & $-$0.111856 & 0.003584 & $-$0.002609 & 0.021888 & 0.007974 & 0.154523 & $-$0.089968 & $-$0.095207 \\
 & z & 1.115700 & 0.030120 & 0.038425 & $-$0.005659 & $-$0.017887 & $-$0.007383 & $-$0.125237 & 0.118998 & 0.031042 \\[3pt]
C2 & x & 0.007077 & 0.338348 & 0.394811 & 0.006015 & 0.024644 & $-$0.028879 & 0.013092 & 0.035083 & $-$0.187826 \\
 & y & $-$0.287281 & $-$0.054442 & 3.350427 & $-$0.004106 & $-$0.002998 & $-$0.046191 & 0.036523 & $-$0.057440 & $-$0.422528 \\
 & z & $-$0.535610 & $-$3.305799 & 0.055489 & 0.013348 & 0.040627 & $-$0.003701 & 0.031497 & 0.461592 & 0.051787 \\[3pt]
Cl3 & x & $-$0.077447 & $-$0.122644 & 2.447633 & 0.003442 & $-$0.023445 & $-$0.006375 & $-$0.074005 & 0.432041 & $-$2.666940 \\
 & y & 0.249569 & 0.148840 & $-$5.508758 & 0.011769 & $-$0.012232 & 0.079407 & $-$0.316792 & 0.136609 & 5.201572 \\
 & z & $-$2.653396 & 5.393630 & $-$0.069531 & 0.025093 & $-$0.073184 & 0.006917 & 2.479895 & $-$5.310478 & $-$0.062614 \\[3pt]
C4 & x & 0.031229 & $-$0.483004 & $-$4.375695 & $-$0.002404 & $-$0.011662 & $-$0.042538 & 0.028825 & 0.080006 & 0.698755 \\
 & y & 0.481369 & 0.008915 & $-$0.263404 & $-$0.006717 & $-$0.001115 & 0.005286 & $-$0.100021 & 0.007800 & 0.068974 \\
 & z & 4.419140 & 0.252509 & $-$0.031466 & 0.049072 & 0.003459 & 0.002108 & $-$0.648776 & $-$0.071123 & $-$0.029358 \\[3pt]
H5 & x & $-$0.054001 & 0.613134 & $-$0.247227 & 0.006167 & $-$0.006291 & 0.005187 & $-$0.047833 & $-$0.113050 & $-$0.000522 \\
 & y & $-$0.584593 & 0.049586 & $-$0.006371 & 0.001111 & $-$0.008571 & 0.000289 & 0.136411 & 0.041015 & 0.021524 \\
 & z & 0.253159 & 0.009244 & $-$0.000783 & $-$0.000501 & $-$0.009057 & 0.002911 & 0.011139 & $-$0.027421 & 0.002127 \\[3pt]
C6 & x & 0.035734 & $-$0.234372 & 1.536746 & $-$0.003263 & $-$0.020220 & 0.007335 & 0.032471 & 0.175825 & $-$0.383851 \\
 & y & 0.367875 & $-$0.042448 & 5.706472 & 0.004608 & 0.000186 & $-$0.019256 & $-$0.057934 & $-$0.042262 & $-$1.069666 \\
 & z & $-$1.647250 & $-$5.690448 & 0.003137 & 0.000988 & 0.019955 & 0.005345 & 0.281669 & 1.086388 & 0.008483 \\[3pt]
H7 & x & 0.001224 & $-$0.147418 & 0.414142 & 0.000175 & $-$0.000511 & 0.004986 & 0.001400 & 0.036008 & $-$0.109453 \\
 & y & 0.142736 & $-$0.011378 & 1.207892 & 0.002755 & 0.002344 & 0.019829 & $-$0.038445 & $-$0.009034 & $-$0.341048 \\
 & z & $-$0.403861 & $-$1.201765 & 0.008642 & $-$0.007245 & $-$0.021181 & $-$0.002804 & 0.117475 & 0.345822 & 0.005837 \\[3pt]
H8 & x & 0.089022 & $-$0.006608 & $-$1.161857 & $-$0.007495 & $-$0.000812 & 0.026951 & 0.081526 & $-$0.071872 & 0.025875 \\
 & y & 0.017145 & $-$0.004776 & 0.107419 & $-$0.004170 & 0.000717 & 0.012566 & 0.077427 & $-$0.004059 & $-$0.205878 \\
 & z & 1.058987 & $-$0.241209 & $-$0.084913 & $-$0.013560 & 0.005905 & 0.005980 & $-$0.115355 & 0.090560 & $-$0.078933 \\[3pt]
H9 & x & $-$0.070082 & $-$0.041308 & $-$1.117900 & 0.006808 & 0.000474 & 0.034115 & $-$0.063274 & $-$0.075302 & $-$0.023752 \\
 & y & 0.004454 & $-$0.001770 & $-$0.303798 & $-$0.000193 & $-$0.001194 & $-$0.003483 & 0.038729 & $-$0.002964 & 0.199262 \\
 & z & 0.977719 & 0.396534 & 0.075582 & $-$0.010635 & $-$0.007276 & $-$0.007019 & $-$0.092949 & $-$0.117285 & 0.068563 \\[3pt]
H10 & x & $-$0.012605 & $-$0.437714 & $-$0.753535 & 0.000760 & 0.007634 & 0.006963 & $-$0.011844 & 0.177456 & 0.064565 \\
 & y & 0.467029 & 0.006097 & $-$0.026223 & $-$0.007871 & 0.001273 & 0.000305 & $-$0.148378 & 0.007370 & 0.003894 \\
 & z & 0.982862 & 0.053373 & 0.010517 & $-$0.034484 & $-$0.001656 & $-$0.001142 & 0.137242 & 0.021904 & 0.009375 \\
\end{tabular}
\end{ruledtabular}
\end{table*}

%% file: si/si_vgapt.tex
\begin{table*}
\caption{CCSD(FC)/cc-pVTZ velocity-gauge atomic polar tensors (VG-APTs) of \cby. The total includes the nuclear contribution (see text). Rows: nuclear displacement $R_{\lambda\alpha}$; columns: dipole component $\gamma$, i.e., $P^{\lambda}_{\alpha\gamma}$ in atomic units ($e$).}
\label{tab:si_vgapt}
\begin{ruledtabular}
\begin{tabular}{llrrr rrr rrr}
 &  & \multicolumn{3}{c}{Reference (HF)} & \multicolumn{3}{c}{Correlation (CCSD)} & \multicolumn{3}{c}{Total} \\
\cline{3-5}\cline{6-8}\cline{9-11}
Atom & $\alpha$ & \multicolumn{1}{c}{$\gamma=$x} & \multicolumn{1}{c}{$\gamma=$y} & \multicolumn{1}{c}{$\gamma=$z} & \multicolumn{1}{c}{$\gamma=$x} & \multicolumn{1}{c}{$\gamma=$y} & \multicolumn{1}{c}{$\gamma=$z} & \multicolumn{1}{c}{$\gamma=$x} & \multicolumn{1}{c}{$\gamma=$y} & \multicolumn{1}{c}{$\gamma=$z} \\
\hline
C1 & x & $-$4.100003 & 0.258675 & 0.078484 & $-$0.160132 & $-$0.064590 & $-$0.021225 & 1.739865 & 0.194085 & 0.057259 \\
 & y & 0.341698 & $-$4.558323 & 0.088642 & $-$0.082788 & $-$0.035826 & $-$0.033242 & 0.258910 & 1.405852 & 0.055400 \\
 & z & 0.142625 & 0.142456 & $-$4.858896 & $-$0.038934 & $-$0.051581 & $-$0.028507 & 0.103691 & 0.090875 & 1.112596 \\[3pt]
C2 & x & $-$5.201038 & 0.076283 & 0.049295 & $-$0.009021 & 0.017069 & 0.002813 & 0.789941 & 0.093352 & 0.052108 \\
 & y & 0.105796 & $-$5.101784 & $-$0.035898 & 0.025720 & 0.041206 & 0.007249 & 0.131516 & 0.939422 & $-$0.028649 \\
 & z & 0.040888 & $-$0.033524 & $-$4.961571 & 0.011955 & 0.006552 & 0.033415 & 0.052842 & $-$0.026972 & 1.071844 \\[3pt]
Cl3 & x & $-$8.942752 & $-$0.400186 & $-$0.143738 & 0.238667 & 0.042096 & 0.015528 & 8.295914 & $-$0.358090 & $-$0.128210 \\
 & y & $-$0.479906 & $-$8.646942 & $-$0.130180 & 0.058189 & 0.187819 & 0.014734 & $-$0.421716 & 8.540877 & $-$0.115446 \\
 & z & $-$0.220985 & $-$0.153758 & $-$8.220346 & 0.027808 & 0.019149 & 0.157542 & $-$0.193177 & $-$0.134609 & 8.937196 \\[3pt]
C4 & x & $-$5.093690 & $-$0.032468 & 0.002964 & $-$0.013256 & 0.007378 & 0.002665 & 0.893054 & $-$0.025090 & 0.005629 \\
 & y & $-$0.054032 & $-$5.001717 & 0.034770 & 0.008879 & $-$0.055517 & $-$0.002768 & $-$0.045153 & 0.942766 & 0.032002 \\
 & z & $-$0.001536 & $-$0.029165 & $-$4.985672 & 0.003100 & 0.012015 & $-$0.024023 & 0.001564 & $-$0.017149 & 0.990305 \\[3pt]
H5 & x & $-$1.054534 & $-$0.015966 & $-$0.009827 & 0.024319 & 0.002878 & 0.000802 & $-$0.030215 & $-$0.013088 & $-$0.009025 \\
 & y & $-$0.030504 & $-$0.987274 & $-$0.011847 & 0.005056 & 0.010690 & 0.000132 & $-$0.025448 & 0.023416 & $-$0.011716 \\
 & z & $-$0.027577 & $-$0.031644 & $-$1.046369 & 0.007946 & 0.004314 & 0.015041 & $-$0.019630 & $-$0.027331 & $-$0.031329 \\[3pt]
C6 & x & $-$5.536312 & 0.234201 & 0.106424 & 0.101080 & $-$0.017154 & $-$0.005540 & 0.564768 & 0.217046 & 0.100885 \\
 & y & 0.264711 & $-$5.036404 & $-$0.028797 & $-$0.030431 & 0.039285 & $-$0.004074 & 0.234280 & 1.002881 & $-$0.032872 \\
 & z & 0.146431 & $-$0.038210 & $-$4.965285 & $-$0.016063 & $-$0.002717 & 0.037501 & 0.130369 & $-$0.040927 & 1.072217 \\[3pt]
H7 & x & $-$0.749535 & $-$0.000148 & 0.002639 & $-$0.017441 & 0.000911 & 0.000548 & 0.233023 & 0.000763 & 0.003188 \\
 & y & 0.005186 & $-$0.756674 & $-$0.005989 & 0.001040 & $-$0.015168 & 0.001511 & 0.006227 & 0.228158 & $-$0.004478 \\
 & z & 0.002148 & $-$0.004549 & $-$0.755723 & 0.000800 & 0.001341 & $-$0.016324 & 0.002947 & $-$0.003208 & 0.227953 \\[3pt]
H8 & x & $-$1.019080 & $-$0.112210 & $-$0.057948 & 0.023348 & 0.012500 & 0.007951 & 0.004268 & $-$0.099709 & $-$0.049997 \\
 & y & $-$0.119747 & $-$1.035113 & $-$0.024372 & 0.017249 & 0.026984 & 0.006105 & $-$0.102498 & $-$0.008130 & $-$0.018267 \\
 & z & $-$0.045399 & $-$0.022128 & $-$0.927320 & 0.004512 & 0.003076 & 0.011922 & $-$0.040887 & $-$0.019051 & 0.084603 \\[3pt]
H9 & x & $-$1.082890 & 0.110892 & 0.051724 & 0.032887 & $-$0.013836 & $-$0.007835 & $-$0.050003 & 0.097055 & 0.043889 \\
 & y & 0.089119 & $-$0.980383 & $-$0.004223 & $-$0.014785 & 0.020817 & 0.003331 & 0.074335 & 0.040434 & $-$0.000892 \\
 & z & 0.047933 & $-$0.006334 & $-$0.924168 & $-$0.006049 & 0.003832 & 0.010433 & 0.041884 & $-$0.002502 & 0.086265 \\[3pt]
H10 & x & $-$0.898697 & 0.007411 & 0.013528 & 0.009077 & 0.001048 & $-$0.001776 & 0.110380 & 0.008459 & 0.011753 \\
 & y & 0.004163 & $-$0.938541 & $-$0.043948 & $-$0.000220 & 0.014404 & 0.000745 & 0.003942 & 0.075863 & $-$0.043203 \\
 & z & 0.009017 & 0.015013 & $-$1.099622 & $-$0.001506 & $-$0.002132 & 0.033881 & 0.007511 & 0.012881 & $-$0.065741 \\
\end{tabular}
\end{ruledtabular}
\end{table*}

%% file: si/si_augref.tex
\begin{table*}
\caption{HF/aug-cc-pCVTZ (all-electron) reference contributions to the LG-APT, AAT, and VG-APT of \cby\ at the CCSD(FC)/cc-pVTZ structure, used in Eq.~(\ref{eq:si_composite}). Rows: nuclear displacement; columns: dipole (APTs) or magnetic-field (AAT) component; atomic units; AAT gauge origin at the center of mass. Nuclear contributions are not included.}
\label{tab:si_augref}
\begin{ruledtabular}
\begin{tabular}{llrrr rrr rrr}
 &  & \multicolumn{3}{c}{LG-APT} & \multicolumn{3}{c}{AAT} & \multicolumn{3}{c}{VG-APT} \\
\cline{3-5}\cline{6-8}\cline{9-11}
Atom & $\alpha$ & \multicolumn{1}{c}{x} & \multicolumn{1}{c}{y} & \multicolumn{1}{c}{z} & \multicolumn{1}{c}{x} & \multicolumn{1}{c}{y} & \multicolumn{1}{c}{z} & \multicolumn{1}{c}{x} & \multicolumn{1}{c}{y} & \multicolumn{1}{c}{z} \\
\hline
C1 & x & $-$4.983834 & 0.239793 & 0.070828 & 0.056863 & 1.132462 & $-$1.055026 & $-$4.917264 & 0.242596 & 0.072193 \\
 & y & 0.325582 & $-$5.452543 & 0.066104 & $-$1.283785 & $-$0.112895 & 0.025542 & 0.328029 & $-$5.385778 & 0.066974 \\
 & z & 0.132706 & 0.135156 & $-$5.741813 & 1.289543 & 0.011335 & 0.039218 & 0.133948 & 0.135928 & $-$5.674056 \\[3pt]
C2 & x & $-$6.040780 & $-$0.090058 & $-$0.034942 & $-$0.009709 & 0.314413 & 0.596304 & $-$5.976574 & $-$0.081694 & $-$0.030731 \\
 & y & $-$0.073863 & $-$6.229035 & $-$0.005634 & $-$0.329387 & $-$0.030042 & 4.091907 & $-$0.065676 & $-$6.151572 & $-$0.006975 \\
 & z & $-$0.058028 & $-$0.000802 & $-$6.167230 & $-$0.672398 & $-$4.089354 & 0.047978 & $-$0.053829 & $-$0.002137 & $-$6.087336 \\[3pt]
Cl3 & x & $-$17.618504 & $-$0.188316 & $-$0.034395 & $-$0.049919 & $-$0.471530 & 5.028937 & $-$17.083214 & $-$0.202733 & $-$0.040549 \\
 & y & $-$0.257929 & $-$17.382094 & $-$0.035214 & 0.559474 & 0.103887 & $-$10.740575 & $-$0.272188 & $-$16.841131 & $-$0.039411 \\
 & z & $-$0.113961 & $-$0.064910 & $-$17.183598 & $-$5.202391 & 10.708388 & $-$0.055801 & $-$0.119892 & $-$0.069050 & $-$16.635167 \\[3pt]
C4 & x & $-$6.007830 & $-$0.034092 & 0.001133 & 0.032036 & $-$0.564176 & $-$5.101978 & $-$5.939313 & $-$0.033898 & 0.001080 \\
 & y & $-$0.052420 & $-$5.918729 & 0.040359 & 0.571586 & 0.011666 & $-$0.309419 & $-$0.051908 & $-$5.852819 & 0.041832 \\
 & z & $-$0.003864 & $-$0.032529 & $-$5.895847 & 5.150509 & 0.301213 & $-$0.034656 & $-$0.003938 & $-$0.031323 & $-$5.827828 \\[3pt]
H5 & x & $-$1.077168 & $-$0.013199 & $-$0.012003 & $-$0.053301 & 0.619812 & $-$0.252968 & $-$1.074596 & $-$0.013431 & $-$0.012058 \\
 & y & $-$0.027641 & $-$1.005822 & $-$0.007359 & $-$0.594207 & 0.049204 & $-$0.009678 & $-$0.027867 & $-$1.003298 & $-$0.007554 \\
 & z & $-$0.025474 & $-$0.027979 & $-$1.033534 & 0.257144 & 0.011167 & $-$0.001021 & $-$0.025611 & $-$0.028074 & $-$1.031582 \\[3pt]
C6 & x & $-$6.471711 & 0.079939 & 0.005086 & 0.016436 & $-$0.389019 & 1.955080 & $-$6.405708 & 0.086259 & 0.008681 \\
 & y & 0.106750 & $-$6.226392 & 0.013593 & 0.448115 & $-$0.014089 & 6.984066 & 0.113331 & $-$6.149525 & 0.012053 \\
 & z & 0.059323 & $-$0.003208 & $-$6.213691 & $-$2.003094 & $-$6.980595 & $-$0.001592 & 0.062868 & $-$0.004728 & $-$6.134827 \\[3pt]
H7 & x & $-$0.754466 & $-$0.006316 & $-$0.001135 & 0.000481 & $-$0.151084 & 0.421675 & $-$0.753283 & $-$0.005862 & $-$0.000879 \\
 & y & 0.000372 & $-$0.771459 & $-$0.004602 & 0.144493 & $-$0.009850 & 1.223255 & 0.000797 & $-$0.769599 & $-$0.004716 \\
 & z & $-$0.000366 & $-$0.003187 & $-$0.770704 & $-$0.409511 & $-$1.217051 & 0.007823 & $-$0.000122 & $-$0.003303 & $-$0.768671 \\[3pt]
H8 & x & $-$1.031077 & $-$0.097646 & $-$0.053111 & 0.083172 & $-$0.002915 & $-$1.178799 & $-$1.028868 & $-$0.097968 & $-$0.053244 \\
 & y & $-$0.101900 & $-$1.048689 & $-$0.016291 & 0.010027 & $-$0.003859 & 0.122791 & $-$0.102311 & $-$1.046391 & $-$0.016530 \\
 & z & $-$0.040842 & $-$0.015236 & $-$0.952189 & 1.083962 & $-$0.249207 & $-$0.079735 & $-$0.040951 & $-$0.015393 & $-$0.949676 \\[3pt]
H9 & x & $-$1.086253 & 0.100626 & 0.046130 & $-$0.063992 & $-$0.037312 & $-$1.129134 & $-$1.084200 & 0.100818 & 0.046313 \\
 & y & 0.075137 & $-$0.996852 & $-$0.001029 & 0.003236 & $-$0.000660 & $-$0.319508 & 0.075272 & $-$0.994306 & $-$0.001140 \\
 & z & 0.042839 & $-$0.002948 & $-$0.949662 & 1.002021 & 0.409655 & 0.068280 & 0.042999 & $-$0.003045 & $-$0.947091 \\[3pt]
H10 & x & $-$0.928378 & 0.009269 & 0.012410 & $-$0.011740 & $-$0.446003 & $-$0.770568 & $-$0.925787 & 0.009246 & 0.012382 \\
 & y & 0.005912 & $-$0.968384 & $-$0.049925 & 0.473516 & 0.005643 & $-$0.026574 & 0.005853 & $-$0.965761 & $-$0.050044 \\
 & z & 0.007667 & 0.015643 & $-$1.091732 & 0.978946 & 0.053386 & 0.010116 & 0.007718 & 0.015615 & $-$1.089874 \\
\end{tabular}
\end{ruledtabular}
\end{table*}

%% file: si/si_hessian.tex
\squeezetable
\begin{table*}
\caption{CCSD(FC)/cc-pVTZ Cartesian Hessian $\partial^2 E/\partial R_{\lambda\alpha}\partial R_{\kappa\beta}$ of \cby\ ($E_\mathrm{h}/a_0^2$; frame and atom order of Table~\ref{tab:si_geom}; total, including nuclear repulsion). The Hessian is symmetric; only the lower triangle is given, in 4 tables (Tables~\ref{tab:si_hess1}--\ref{tab:si_hess4}). Columns: atoms C1--Cl3.}
\label{tab:si_hess1}
\begin{ruledtabular}
\begin{tabular}{llrrr rrr rrr}
 &  & \multicolumn{3}{c}{C1} & \multicolumn{3}{c}{C2} & \multicolumn{3}{c}{Cl3} \\
\cline{3-5}\cline{6-8}\cline{9-11}
Atom & $\alpha$ & \multicolumn{1}{c}{$\beta=$x} & \multicolumn{1}{c}{$\beta=$y} & \multicolumn{1}{c}{$\beta=$z} & \multicolumn{1}{c}{$\beta=$x} & \multicolumn{1}{c}{$\beta=$y} & \multicolumn{1}{c}{$\beta=$z} & \multicolumn{1}{c}{$\beta=$x} & \multicolumn{1}{c}{$\beta=$y} & \multicolumn{1}{c}{$\beta=$z} \\
\hline
C1 & x & 0.43012605 &  &  &  &  &  &  &  &  \\
 & y & $-$0.08164280 & 0.41421724 &  &  &  &  &  &  &  \\
 & z & $-$0.05009085 & $-$0.00854893 & 0.55901956 &  &  &  &  &  &  \\[3pt]
C2 & x & $-$0.23048896 & 0.09548301 & 0.05147801 & 1.14154206 &  &  &  &  &  \\
 & y & 0.08606992 & $-$0.10368406 & $-$0.02165202 & $-$0.48809648 & 0.32225812 &  &  &  &  \\
 & z & 0.04786629 & $-$0.02110409 & $-$0.08205570 & $-$0.26530750 & 0.12218264 & 0.15670597 &  &  &  \\[3pt]
Cl3 & x & $-$0.08026719 & $-$0.04358112 & $-$0.02456650 & $-$0.03694213 & $-$0.01704608 & $-$0.00872513 & 0.13086333 &  &  \\
 & y & $-$0.03573457 & $-$0.06495252 & $-$0.01929138 & $-$0.00402082 & 0.00453032 & 0.00224224 & 0.07864742 & 0.09335290 &  \\
 & z & $-$0.02132803 & $-$0.01951533 & $-$0.04623180 & $-$0.00111932 & 0.00195643 & 0.00341490 & 0.03977189 & 0.03034975 & 0.04037260 \\[3pt]
C4 & x & $-$0.07360278 & 0.02063181 & $-$0.00458763 & $-$0.00653001 & 0.03189848 & $-$0.01288516 & $-$0.00891026 & $-$0.02913011 & 0.00469660 \\
 & y & 0.01651127 & $-$0.17520843 & 0.05432928 & 0.01251711 & $-$0.01954516 & 0.00483569 & $-$0.00901155 & $-$0.02351428 & 0.00407629 \\
 & z & $-$0.00250199 & 0.05417340 & $-$0.10839169 & $-$0.00021502 & $-$0.00777997 & 0.00553621 & $-$0.00428777 & $-$0.01231709 & 0.00365557 \\[3pt]
H5 & x & $-$0.04597144 & 0.00750253 & 0.02766528 & $-$0.00711668 & 0.00558963 & 0.03179006 & $-$0.00941890 & $-$0.01127094 & $-$0.02331314 \\
 & y & 0.00760075 & $-$0.04647289 & $-$0.01889280 & 0.00545535 & $-$0.00117184 & $-$0.01645416 & $-$0.00942500 & $-$0.00784204 & $-$0.01789637 \\
 & z & 0.02775070 & $-$0.01636435 & $-$0.32289975 & 0.00526527 & $-$0.00172780 & $-$0.00550314 & $-$0.00376191 & $-$0.00308361 & $-$0.00338852 \\[3pt]
C6 & x & 0.00240951 & $-$0.00016820 & 0.00080861 & $-$0.86780615 & 0.38280674 & 0.20931157 & 0.00061326 & 0.00095598 & 0.00042253 \\
 & y & 0.00046503 & 0.00264337 & $-$0.00119292 & 0.38357969 & $-$0.20256746 & $-$0.09607062 & $-$0.00038180 & 0.00021088 & 0.00000207 \\
 & z & 0.00077367 & $-$0.00109422 & 0.00313634 & 0.21025061 & $-$0.09633806 & $-$0.07986075 & $-$0.00020725 & $-$0.00024169 & 0.00037013 \\[3pt]
H7 & x & 0.00083809 & 0.00186316 & 0.00113137 & 0.00648195 & $-$0.00198276 & $-$0.00120424 & 0.00088453 & $-$0.00002133 & $-$0.00000858 \\
 & y & 0.00202000 & 0.00471167 & $-$0.00073572 & $-$0.00286672 & 0.00258772 & 0.00126860 & 0.00095883 & $-$0.00006606 & $-$0.00003026 \\
 & z & 0.00089944 & $-$0.00066154 & 0.00604982 & $-$0.00109767 & 0.00102699 & 0.00114589 & 0.00053857 & $-$0.00000056 & $-$0.00009759 \\[3pt]
H8 & x & 0.00099552 & 0.00282846 & 0.00109772 & 0.00021943 & 0.00050041 & 0.00042875 & 0.00176646 & 0.00068557 & 0.00015245 \\
 & y & $-$0.02129410 & $-$0.01699617 & $-$0.01051140 & $-$0.00076970 & 0.00103699 & $-$0.00020141 & $-$0.00166951 & $-$0.00563137 & 0.00165265 \\
 & z & 0.01062229 & 0.00811998 & 0.00334318 & 0.00012283 & $-$0.00035170 & 0.00030666 & 0.00110595 & 0.00344422 & 0.00070564 \\[3pt]
H9 & x & $-$0.00296124 & $-$0.00192307 & 0.00094837 & 0.00023924 & 0.00075785 & $-$0.00034161 & 0.00133996 & $-$0.00043131 & 0.00009340 \\
 & y & 0.02774190 & $-$0.01408036 & $-$0.00849141 & 0.00044385 & $-$0.00525587 & 0.00320936 & 0.00029216 & 0.00176528 & $-$0.00025633 \\
 & z & $-$0.01437549 & 0.00686422 & 0.00311398 & 0.00022289 & 0.00313831 & $-$0.00041854 & $-$0.00022364 & $-$0.00073542 & $-$0.00016400 \\[3pt]
H10 & x & $-$0.00107758 & $-$0.00099378 & $-$0.00388435 & 0.00040127 & $-$0.00049772 & $-$0.00093303 & 0.00007094 & 0.00032011 & 0.00063220 \\
 & y & $-$0.00173741 & $-$0.00017786 & 0.03498731 & $-$0.00172528 & 0.00181124 & 0.00009175 & 0.00121666 & 0.00214688 & $-$0.00033889 \\
 & z & 0.00038398 & $-$0.00186914 & $-$0.01508394 & 0.00039991 & $-$0.00045484 & 0.00072851 & 0.00035580 & $-$0.00036646 & 0.00136308 \\
\end{tabular}
\end{ruledtabular}
\end{table*}

\squeezetable
\begin{table*}
\caption{CCSD(FC)/cc-pVTZ Cartesian Hessian $\partial^2 E/\partial R_{\lambda\alpha}\partial R_{\kappa\beta}$ of \cby\ ($E_\mathrm{h}/a_0^2$; frame and atom order of Table~\ref{tab:si_geom}; total, including nuclear repulsion). Continuation of Table~\ref{tab:si_hess1}. Columns: atoms C4--C6.}
\label{tab:si_hess2}
\begin{ruledtabular}
\begin{tabular}{llrrr rrr rrr}
 &  & \multicolumn{3}{c}{C4} & \multicolumn{3}{c}{H5} & \multicolumn{3}{c}{C6} \\
\cline{3-5}\cline{6-8}\cline{9-11}
Atom & $\alpha$ & \multicolumn{1}{c}{$\beta=$x} & \multicolumn{1}{c}{$\beta=$y} & \multicolumn{1}{c}{$\beta=$z} & \multicolumn{1}{c}{$\beta=$x} & \multicolumn{1}{c}{$\beta=$y} & \multicolumn{1}{c}{$\beta=$z} & \multicolumn{1}{c}{$\beta=$x} & \multicolumn{1}{c}{$\beta=$y} & \multicolumn{1}{c}{$\beta=$z} \\
\hline
C4 & x & 0.59600328 &  &  &  &  &  &  &  &  \\
 & y & 0.00136989 & 0.49891084 &  &  &  &  &  &  &  \\
 & z & $-$0.00448408 & 0.03905576 & 0.57591062 &  &  &  &  &  &  \\[3pt]
H5 & x & 0.00209225 & 0.00005376 & $-$0.00374813 & 0.06106405 &  &  &  &  &  \\
 & y & $-$0.00220929 & 0.00160898 & 0.03283252 & $-$0.00299914 & 0.05740249 &  &  &  &  \\
 & z & 0.00154696 & $-$0.00005309 & $-$0.01205950 & $-$0.03001704 & 0.01860982 & 0.34200631 &  &  &  \\[3pt]
C6 & x & $-$0.00164767 & $-$0.00273734 & 0.00197332 & $-$0.00140292 & 0.00088842 & $-$0.00067897 & 1.20251486 &  &  \\
 & y & 0.00024742 & 0.00087089 & $-$0.00046235 & 0.00080780 & $-$0.00017318 & 0.00035642 & $-$0.53076867 & 0.28957870 &  \\
 & z & 0.00087543 & 0.00109420 & $-$0.00052559 & $-$0.00062546 & 0.00036837 & 0.00042124 & $-$0.29034117 & 0.13492159 & 0.11710969 \\[3pt]
H7 & x & $-$0.00020630 & 0.00021071 & 0.00004296 & $-$0.00018715 & 0.00012412 & 0.00008765 & $-$0.33531808 & 0.14617047 & 0.07927012 \\
 & y & $-$0.00101640 & 0.00033585 & 0.00034511 & 0.00000514 & $-$0.00005317 & $-$0.00009574 & 0.14723524 & $-$0.09048178 & $-$0.03797171 \\
 & z & 0.00057528 & $-$0.00033889 & $-$0.00021321 & $-$0.00106849 & 0.00055242 & 0.00012061 & 0.07943540 & $-$0.03779432 & $-$0.04120759 \\[3pt]
H8 & x & $-$0.19637970 & $-$0.11845144 & $-$0.06295454 & 0.00014375 & $-$0.00033834 & 0.00037197 & 0.00026070 & $-$0.00010099 & $-$0.00007929 \\
 & y & $-$0.11591320 & $-$0.14308242 & $-$0.04929078 & 0.00007144 & 0.00119309 & 0.00020201 & 0.00053147 & 0.00020141 & $-$0.00003230 \\
 & z & $-$0.06341625 & $-$0.05043186 & $-$0.07590146 & 0.00026882 & $-$0.00087109 & 0.00025717 & 0.00005059 & 0.00002788 & $-$0.00011639 \\[3pt]
H9 & x & $-$0.26119785 & 0.09737097 & 0.06214062 & 0.00006922 & 0.00012911 & $-$0.00041093 & 0.00040555 & 0.00011599 & $-$0.00013279 \\
 & y & 0.09231626 & $-$0.09042288 & $-$0.02756739 & $-$0.00056172 & 0.00129059 & $-$0.00034591 & 0.00098246 & $-$0.00022310 & $-$0.00029449 \\
 & z & 0.06399637 & $-$0.02893774 & $-$0.06855034 & $-$0.00035911 & $-$0.00086794 & 0.00057551 & $-$0.00063345 & 0.00015580 & 0.00038849 \\[3pt]
H10 & x & $-$0.04962096 & 0.00216663 & 0.01403464 & 0.00072781 & 0.00077401 & $-$0.00015368 & $-$0.00002905 & $-$0.00013494 & 0.00021614 \\
 & y & 0.00180516 & $-$0.04995338 & $-$0.02898920 & 0.00080149 & $-$0.00578205 & 0.00250224 & 0.00027391 & $-$0.00005972 & $-$0.00041168 \\
 & z & 0.01368249 & $-$0.02362964 & $-$0.31946059 & $-$0.00059279 & 0.00261923 & 0.00047008 & $-$0.00034843 & 0.00005646 & 0.00028443 \\
\end{tabular}
\end{ruledtabular}
\end{table*}

\squeezetable
\begin{table*}
\caption{CCSD(FC)/cc-pVTZ Cartesian Hessian $\partial^2 E/\partial R_{\lambda\alpha}\partial R_{\kappa\beta}$ of \cby\ ($E_\mathrm{h}/a_0^2$; frame and atom order of Table~\ref{tab:si_geom}; total, including nuclear repulsion). Continuation of Table~\ref{tab:si_hess1}. Columns: atoms H7--H9.}
\label{tab:si_hess3}
\begin{ruledtabular}
\begin{tabular}{llrrr rrr rrr}
 &  & \multicolumn{3}{c}{H7} & \multicolumn{3}{c}{H8} & \multicolumn{3}{c}{H9} \\
\cline{3-5}\cline{6-8}\cline{9-11}
Atom & $\alpha$ & \multicolumn{1}{c}{$\beta=$x} & \multicolumn{1}{c}{$\beta=$y} & \multicolumn{1}{c}{$\beta=$z} & \multicolumn{1}{c}{$\beta=$x} & \multicolumn{1}{c}{$\beta=$y} & \multicolumn{1}{c}{$\beta=$z} & \multicolumn{1}{c}{$\beta=$x} & \multicolumn{1}{c}{$\beta=$y} & \multicolumn{1}{c}{$\beta=$z} \\
\hline
H7 & x & 0.32747661 &  &  &  &  &  &  &  &  \\
 & y & $-$0.14638176 & 0.08283959 &  &  &  &  &  &  &  \\
 & z & $-$0.07927014 & 0.03734439 & 0.03422331 &  &  &  &  &  &  \\[3pt]
H8 & x & $-$0.00000046 & 0.00003176 & $-$0.00001305 & 0.20984415 &  &  &  &  &  \\
 & y & 0.00000613 & $-$0.00008176 & $-$0.00002398 & 0.12986519 & 0.15571652 &  &  &  &  \\
 & z & 0.00001402 & $-$0.00000773 & 0.00000556 & 0.06947778 & 0.05321612 & 0.07728955 &  &  &  \\[3pt]
H9 & x & 0.00004059 & $-$0.00003975 & $-$0.00000526 & $-$0.01841463 & 0.01002398 & 0.00656186 & 0.28059147 &  &  \\
 & y & 0.00000613 & 0.00023778 & $-$0.00014662 & $-$0.01753055 & 0.00919987 & 0.00632442 & $-$0.10520218 & 0.09915545 &  \\
 & z & $-$0.00004018 & $-$0.00013572 & 0.00002152 & $-$0.00964595 & 0.00536371 & 0.00198237 & $-$0.06822878 & 0.02801120 & 0.06912232 \\[3pt]
H10 & x & $-$0.00000977 & 0.00005367 & 0.00000592 & 0.00156477 & $-$0.00085170 & $-$0.02480790 & $-$0.00011231 & 0.00151170 & 0.02928734 \\
 & y & 0.00000512 & $-$0.00002985 & 0.00004212 & 0.00250993 & $-$0.00155616 & $-$0.01947024 & $-$0.00080158 & $-$0.00166675 & $-$0.01285643 \\
 & z & $-$0.00002297 & 0.00001878 & $-$0.00004832 & 0.00116416 & $-$0.00037460 & $-$0.00787228 & $-$0.00062487 & $-$0.00044283 & $-$0.00607130 \\
\end{tabular}
\end{ruledtabular}
\end{table*}

\squeezetable
\begin{table*}
\caption{CCSD(FC)/cc-pVTZ Cartesian Hessian $\partial^2 E/\partial R_{\lambda\alpha}\partial R_{\kappa\beta}$ of \cby\ ($E_\mathrm{h}/a_0^2$; frame and atom order of Table~\ref{tab:si_geom}; total, including nuclear repulsion). Continuation of Table~\ref{tab:si_hess1}. Columns: atoms H10.}
\label{tab:si_hess4}
\begin{ruledtabular}
\begin{tabular}{llrrr}
 &  & \multicolumn{3}{c}{H10} \\
\cline{3-5}
Atom & $\alpha$ & \multicolumn{1}{c}{$\beta=$x} & \multicolumn{1}{c}{$\beta=$y} & \multicolumn{1}{c}{$\beta=$z} \\
\hline
H10 & x & 0.04808488 &  &  \\
 & y & $-$0.00234799 & 0.05526764 &  \\
 & z & $-$0.01439728 & 0.02444304 & 0.34569034 \\
\end{tabular}
\end{ruledtabular}
\end{table*}

%% file: references.bib
@article{Stephens85,
    author = {P. J. Stephens},
    journal = {J. Phys. Chem.},
    pages = {748-752},
    title = {Theory of Vibrational Circular Dichroism},
    volume = 89,
    year = 1985}

@misc{Krishnan2026:density,
    author = {A. Krishnan and B. M. Shumberger and T. D. Crawford},
    journal = {J. Chem. Phys.},
    title = {A Density-Matrix Formulation of Vibrational Circular Dichroism Spectra for {M{\o}ller-Plesset} and Configuration Interaction Methods},
    year = {2026},
    note = {submitted; arXiv:2610.35392 [physics.chem-ph]}}

@article{Gauss2006,
	author = {J. Gauss and A. Tajti and M. K{\'a}llay and J. F. Stanton and P. G. Szalay},
	journal = {J. Chem. Phys.},
	pages = {144111},
	title = {Analytic calculation of the diagonal Born-Oppenheimer correction within configuration-interaction and coupled-cluster theory},
	volume = {125},
	year = {2006}}

@ARTICLE{Amos1987,
  author = {R. D. Amos and N. C. Handy and K. J. Jalkanen
    and P.J. Stephens},
  title = {Efficient calculation of vibrational magnetic
    dipole transition moments and rotational strengths},
  journal = {Chem. Phys. Lett.},
  year = {1987},
  volume = {133},
  pages = {21--26},
  number = {1},
  doi = {10.1016/0009-2614(87)80046-5},
}

@ARTICLE{Amos1988,
  author = {Roger D. Amos and Karl J. Jalkanen and Philip J. Stephens},
  title = {Alternative Formalism for the Calculation of Atomic Polar Tensors and Atomic 
    Axial Tensors},
  journal = {J. Phys. Chem.},
  year = {1988},
  volume = {92},
  pages = {5571--5575},
  number = {20},
  doi = {},
}

@article{Matthews2020:CFOUR,
	author = {Matthews, Devin A. and Cheng, Lan and Harding, Michael E. and Lipparini, Filippo and Stopkowicz, Stella and Jagau, Thomas-C. and Szalay, P{\'e}ter G. and Gauss, J{\"u}rgen and Stanton, John F.},
	journal = {J. Chem. Phys.},
	month = {06},
	number = {21},
	pages = {214108},
	title = {Coupled-cluster techniques for computational chemistry: The {CFOUR} program package},
	volume = {152},
	year = {2020}}

@article{Dunning89,
	author = {Dunning, Jr., T. H.},
	journal = {J. Chem. Phys.},
	number = 2,
	pages = {1007-1023},
	title = {Gaussian basis sets for use in correlated molecular calculations. {I}. {T}he atoms boron through neon},
	volume = 90,
	year = 1989}

@article{Woon93,
	author = {D. E. Woon and Dunning, Jr., T. H.},
	journal = {J. Chem. Phys.},
	number = 2,
	pages = {1358-1371},
	title = {Gaussian basis sets for use in correlated molecular calculations. {III}. {T}he atoms aluminum through argon},
	volume = 98,
	year = 1993}

@article{pycc,
  author={T. D. Crawford and B. G. Peyton and Z. Wang and A. Krishnan and
  B. M. Shumberger and J. Madriaga and J. Atta-Kumi and A. {De Clercq} and R. Prabhu},
  title={{PyCC}: A Reference Quantum Chemistry {Python} Package for Education, Validation, and Rapid Prototyping},
  journal={arXiv:2610.04107 [physics.chem-ph]},
  year={2026},
  note={\url{https://github.com/CrawfordGroup/pycc}}
}

@article{Smith20,
	author = {Smith,Daniel G. A. and Burns,Lori A. and Simmonett,Andrew C. and Parrish,Robert M. and Schieber,Matthew C. and Galvelis,Raimondas and Kraus,Peter and Kruse,Holger and Di Remigio,Roberto and Alenaizan,Asem and James,Andrew M. and Lehtola,Susi and Misiewicz,Jonathon P. and Scheurer,Maximilian and Shaw,Robert A. and Schriber,Jeffrey B. and Xie,Yi and Glick,Zachary L. and Sirianni,Dominic A. and O'Brien,Joseph Senan and Waldrop,Jonathan M. and Kumar,Ashutosh and Hohenstein,Edward G. and Pritchard,Benjamin P. and Brooks,Bernard R. and Schaefer,Henry F. and Sokolov,Alexander Yu. and Patkowski,Konrad and DePrince,A. Eugene and Bozkaya,U{\u g}ur and King,Rollin A. and Evangelista,Francesco A. and Turney,Justin M. and Crawford,T. Daniel and Sherrill,C. David},
	journal = {The Journal of Chemical Physics},
	number = {18},
	pages = {184108},
	title = {Psi4 1.4: Open-source software for high-throughput quantum chemistry},
	volume = {152},
	year = {2020}}

@article{Stephens1990,
    author = {Stephens, P. J. and Jalkanen, K. J. and Amos, R. D. and Lazzeretti, P. and Zanasi, R.},
    doi = {10.1021/j100368a020},
    journal = {J. Phys. Chem.},
    number = {5},
    pages = {1811--1830},
    title = {\textit{Ab initio} calculations of atomic polar and axial tensors for hydrogen fluoride, water, ammonia, and methane},
    volume = {94},
    year = {1990}}

@article{Woon95,
	author = {D. E. Woon and T. H. Dunning},
	journal = {J. Chem. Phys.},
	number = 11,
	pages = 4572,
	title = {Gaussian basis sets for use in correlated molecular calculations. V. Core-valence basis sets for boron through neon},
	volume = 103,
	year = 1995}

@article{Kendall92,
	author = {R. A. Kendall and T. H. Dunning, Jr. and R. J. Harrison},
	journal = {J. Chem. Phys.},
	number = 9,
	pages = {6796--6806},
	title = {Electron affinities of the first-row atoms revisited. {S}ystematic basis sets and wave functions},
	volume = 96,
	year = 1992,
	doi = {10.1063/1.462569}}

@article{Peterson2002,
	author = {Peterson, Kirk A. and Dunning, Thom H., Jr.},
	journal = {The Journal of Chemical Physics},
	month = {12},
	number = {23},
	pages = {10548-10560},
	title = {Accurate correlation consistent basis sets for molecular core--valence correlation effects: The second row atoms Al--Ar, and the first row atoms B--Ne revisited},
	volume = {117},
	year = {2002}}

@article{Shumberger2026:LGOI,
	author = {B. M. Shumberger and J. R. Cheeseman and M. Caricato and T. D. Crawford},
	journal = {J. Chem. Phys.},
	pages = {124103},
	title = {A Length-Gauge Origin-Invariant Approach to Vibrational Circular Dichroism Spectra Without Gauge-Including Atomic Orbitals},
	volume = {164},
	year = {2026}}

@article{Shumberger2024,
	author = {B. M. Shumberger and T. D. Crawford},
	journal = {J. Chem. Theory Comput.},
	number = {16},
	pages = {7254-7263},
	title = {{Simulation of Vibrational Circular Dichroism Spectra Using Second-Order M{\o}ller-Plesset Perturbation Theory and Configuration Interaction Doubles}},
	volume = {20},
	year = {2024}}

@article{Shumberger2026,
	author = {B. M. Shumberger and T. D. Crawford},
	journal = {J. Phys. Chem. A},
	number = {4},
	pages = {973-983},
	title = {Analytic Computation of Vibrational Circular Dichroism Spectra Using Configuration Interaction Methods},
	volume = {130},
	year = {2026}}

@article{Buckingham1987,
	author = {A. D. Buckingham and P. W. Fowler and P. A. Galwas},
	journal = {Chem. Phys.},
	pages = {1-14},
	title = {Velocity-Dependent Property Surfaces and the Theory of Vibrational Circular Dichroism},
	volume = 112,
	year = 1987}

@article{Peters2022,
	author = {Peters, Laurens D. M. and Culpitt, Tanner and Tellgren, Erik I. and Helgaker, Trygve},
	journal = {The Journal of Chemical Physics},
	month = {10},
	number = {13},
	pages = {134108},
	title = {Magnetic-translational sum rule and approximate models of the molecular Berry curvature},
	volume = {157},
	year = {2022}}

@article{Steinmetz2025,
	author = {Steinmetz, Dominik and Pausch, Ansgar},
	journal = {The Journal of Chemical Physics},
	month = {04},
	number = {15},
	pages = {154109},
	title = {Novel pseudomomentum-translational sum rule for the molecular Berry curvature},
	volume = {162},
	year = {2025}}

@article{Lazzeretti1986,
	author = {Lazzeretti, P. and Zanasi, R.},
	journal = {Phys. Rev. A},
	month = {Jun},
	pages = {3727--3741},
	title = {Electric and magnetic nuclear shielding tensors: A study of the water molecule},
	volume = {33},
	year = {1986}}

@article{Lazzeretti1985,
	author = {Lazzeretti, Paolo and Zanasi, Riccardo},
	journal = {Phys. Rev. A},
	month = {Nov},
	pages = {2607--2614},
	title = {Quantum-mechanical sum rules and gauge invariance: A study of the HF molecule},
	volume = {32},
	year = {1985}}

@ARTICLE{Schrader1972,
  author = {B. Schrader and E. H. Korte},
  title = {Infrared Rotatory Dispersion ({IRD})},
  journal = {Angew. Chem. Int. Ed. Engl.},
  year = {1972},
  volume = {11},
  pages = {226--227},
  number = {3},
  doi = {10.1002/anie.197202261},
}

@ARTICLE{Dudley1972,
  author = {R. J. Dudley and S. F. Mason and R. D. Peacock},
  title = {Infrared vibrational circular dichroism},
  journal = {J. Chem. Soc., Chem. Commun.},
  year = {1972},
  pages = {1084},
  doi = {10.1039/c39720001084},
}

@ARTICLE{Chabay1972,
  author = {I. Chabay},
  title = {Absorptive and scattering circular dichroism of cholesteric liquid
    crystals in the infrared},
  journal = {Chem. Phys. Lett.},
  year = {1972},
  volume = {17},
  pages = {283--287},
  number = {2},
  doi = {10.1016/0009-2614(72)87075-1},
}

@ARTICLE{Holzwarth1973,
  author = {G. Holzwarth and I. Chabay and N. A. W. Holzwarth},
  title = {Infrared circular dichroism and linear dichroism of liquid crystals},
  journal = {J. Chem. Phys.},
  year = {1973},
  volume = {58},
  pages = {4816--4819},
  number = {11},
  doi = {10.1063/1.1679064},
}

@ARTICLE{Hsu1973,
  author = {E. C. Hsu and G. Holzwarth},
  title = {Vibrational circular dichroism observed in crystalline
    $\alpha$-{NiSO$_4$}$\cdot$6{H$_2$O} and $\alpha$-{ZnSeO$_4$}$\cdot$6{H$_2$O}
    between 1900 and 5000 cm$^{-1}$},
  journal = {J. Chem. Phys.},
  year = {1973},
  volume = {59},
  pages = {4678--4685},
  number = {9},
  doi = {10.1063/1.1680680},
}

@ARTICLE{Holzwarth1974,
  author = {G. Holzwarth and Edward C. Hsu and Harry S.
    Mosher and Thomas R. Faulkner and Albert Moscowitz},
  title = {Infrared circular dichroism of carbon-hydrogen
    and carbon-deuterium stretching modes},
  journal = {J. Am. Chem. Soc.},
  year = {1974},
  volume = {96},
  pages = {251--252},
  number = {1},
  doi = {10.1021/ja00808a042},
}

@article{Merten2024,
	author = {Bloino, Julien and J{\"a}hnigen, Sascha and Merten, Christian},
	journal = {The Journal of Physical Chemistry Letters},
	number = {34},
	pages = {8813-8828},
	title = {After 50 Years of Vibrational Circular Dichroism Spectroscopy: Challenges and Opportunities of Increasingly Accurate and Complex Experiments and Computations},
	volume = {15},
	year = {2024}}

@ARTICLE{Nafie2020,
  author = {L. A. Nafie},
  title = {Vibrational optical activity: From discovery and development to
    future challenges},
  journal = {Chirality},
  year = {2020},
  volume = {32},
  pages = {667--692},
  number = {5},
  doi = {10.1002/chir.23191},
}

@ARTICLE{Krupova2020,
  author = {M. Krupov{\'a} and J. Kessler and P. Bou{\v{r}}},
  title = {Recent Trends in Chiroptical Spectroscopy: Theory and Applications
    of Vibrational Circular Dichroism and {R}aman Optical Activity},
  journal = {ChemPlusChem},
  year = {2020},
  volume = {85},
  pages = {561--575},
  number = {3},
  doi = {10.1002/cplu.202000014},
}

@ARTICLE{Magyarfalvi2011,
  author = {G. Magyarfalvi and G. Tarczay and E. Vass},
  title = {Vibrational circular dichroism},
  journal = {WIREs Comput. Mol. Sci.},
  year = {2011},
  volume = {1},
  pages = {403--425},
  number = {3},
  doi = {10.1002/wcms.39},
}

@article{Nafie97,
	author = {L. A. Nafie},
	journal = {Ann. Rev. Phys. Chem.},
	pages = {357-386},
	title = {Infrared and Raman Vibrational Optical Activity},
	volume = 48,
	year = 1997}

@ARTICLE{Craig1978,
  author = {D. P. Craig and T. Thirunamachandran},
  title = {A theory of vibrational circular dichroism in terms of vibronic
    interactions},
  journal = {Mol. Phys.},
  year = {1978},
  volume = {35},
  pages = {825--840},
  number = {3},
  doi = {10.1080/00268977800100611},
}

@ARTICLE{Nafie1983:VCT,
  author = {L. A. Nafie and T. B. Freedman},
  title = {Vibronic coupling theory of infrared vibrational transitions},
  journal = {J. Chem. Phys.},
  year = {1983},
  volume = {78},
  pages = {7108--7116},
  number = {12},
  doi = {10.1063/1.444741},
}

@ARTICLE{Nafie83,
  author = {L. A. Nafie},
  title = {Adiabatic molecular properties beyond the {B}orn--{O}ppenheimer
    approximation. {C}omplete adiabatic wave functions and vibrationally
    induced electronic current density},
  journal = {J. Chem. Phys.},
  year = {1983},
  volume = {79},
  pages = {4950--4957},
  number = {10},
  doi = {10.1063/1.445588},
}

@ARTICLE{Lowe86:CPL,
  author = {M. A. Lowe and P. J. Stephens and G. A. Segal},
  title = {The theory of vibrational circular dichroism:
    \textit{trans}-1,2-dideuteriocyclobutane and propylene oxide},
  journal = {Chem. Phys. Lett.},
  year = {1986},
  volume = {123},
  pages = {108--116},
  number = {1-2},
  doi = {10.1016/0009-2614(86)87023-3},
}

@ARTICLE{Lowe86:JACS,
  author = {M. A. Lowe and G. A. Segal and P. J. Stephens},
  title = {The theory of vibrational circular dichroism:
    \textit{trans}-1,2-dideuteriocyclopropane},
  journal = {J. Am. Chem. Soc.},
  year = {1986},
  volume = {108},
  pages = {248--256},
  number = {2},
  doi = {10.1021/ja00262a011},
}

@ARTICLE{Bak93,
  author = {K. L. Bak and P. J{\o}rgensen and T. Helgaker and K. Ruud and
    H. J. {Aa}. Jensen},
  title = {Gauge-origin independent multiconfigurational self-consistent-field
    theory for vibrational circular dichroism},
  journal = {J. Chem. Phys.},
  year = {1993},
  volume = {98},
  pages = {8873--8887},
  number = {11},
  doi = {10.1063/1.464445},
}

@ARTICLE{Bak94,
  author = {K. L. Bak and P. J{\o}rgensen and T. Helgaker and K. Ruud},
  title = {Basis set convergence and correlation effects in vibrational
    circular dichroism calculations using {L}ondon atomic orbitals},
  journal = {Faraday Discuss.},
  year = {1994},
  volume = {99},
  pages = {121--129},
  doi = {10.1039/fd9949900121},
}

@ARTICLE{Cheeseman96,
  author = {J. R. Cheeseman and M. J. Frisch and F. J. Devlin and
    P. J. Stephens},
  title = {\textit{Ab initio} calculation of atomic axial tensors and
    vibrational rotational strengths using density functional theory},
  journal = {Chem. Phys. Lett.},
  year = {1996},
  volume = {252},
  pages = {211--220},
  doi = {10.1016/0009-2614(96)00154-6},
}

@ARTICLE{Yang1994,
  author = {D. Yang and A. Rauk},
  title = {Vibrational circular dichroism intensities by \textit{ab initio}
    second-order {M{\o}ller--Plesset} vibronic coupling theory},
  journal = {J. Chem. Phys.},
  year = {1994},
  volume = {100},
  pages = {7995--8002},
  number = {11},
  doi = {10.1063/1.466792},
}

@ARTICLE{Scherrer2013,
  author = {A. Scherrer and R. Vuilleumier and D. Sebastiani},
  title = {Nuclear Velocity Perturbation Theory of Vibrational Circular
    Dichroism},
  journal = {J. Chem. Theory Comput.},
  year = {2013},
  volume = {9},
  pages = {5305--5312},
  number = {12},
  doi = {10.1021/ct400700c},
}

@ARTICLE{Scherrer2016,
  author = {A. Scherrer and R. Vuilleumier and D. Sebastiani},
  title = {Vibrational circular dichroism from \textit{ab initio} molecular
    dynamics and nuclear velocity perturbation theory in the liquid phase},
  journal = {J. Chem. Phys.},
  year = {2016},
  volume = {145},
  pages = {084101},
  number = {8},
  doi = {10.1063/1.4960653},
}

@ARTICLE{Ditler2022,
  author = {E. Ditler and T. Zimmermann and C. Kumar and S. Luber},
  title = {Implementation of Nuclear Velocity Perturbation and Magnetic Field
    Perturbation Theory in {CP2K} and Their Application to Vibrational
    Circular Dichroism},
  journal = {J. Chem. Theory Comput.},
  year = {2022},
  volume = {18},
  pages = {2448--2461},
  number = {4},
  doi = {10.1021/acs.jctc.2c00006},
}

@ARTICLE{Shumberger2025,
  author = {B. M. Shumberger and K. C. Pearce and T. D. Crawford},
  title = {Analytic Computation of Vibrational Circular Dichroism Spectra Using
    Second-Order {M{\o}ller-Plesset} Perturbation Theory},
  journal = {J. Chem. Theory Comput.},
  year = {2025},
  volume = {21},
  pages = {3504--3512},
  number = {7},
  doi = {10.1021/acs.jctc.5c00047},
}

@article{He2004,
    author = {He, Jiangtao and Petrovich, Ana and Polavarapu, Prasad L.},
    journal = {J. Phys. Chem. A},
    month = {02},
    number = {10},
    pages = {1671-1680},
    title = {Quantitative Determination of Conformer Populations: Assessment of Specific Rotation, Vibrational Absorption, and Vibrational Circular Dichroism in Substituted Butynes},
    volume = {108},
    year = {2004}}
